\documentclass[aps,pra,10pt,showkeys,twocolumn]{revtex4-1}

\usepackage{xcolor}
\usepackage[colorlinks,citecolor=ltviolet,urlcolor=darkviolet]{hyperref}
\usepackage{cancel} 
\usepackage{comment}
\usepackage{dcolumn}

\usepackage{graphicx}
\usepackage{amsmath,amssymb,amsfonts,upgreek,wrapfig}
\usepackage[rightcaption]{sidecap}
\usepackage{float}
\usepackage{blindtext}
\usepackage{nicefrac}

\definecolor{uablue}{rgb}{0.0, 0.2, 0.67}
\definecolor{uared}{rgb}{0.85, 0.0, 0.3}
\definecolor{persianblue}{rgb}{0.11, 0.22, 0.73}
\definecolor{persianindigo}{rgb}{0.2, 0.07, 0.48}
\definecolor{persianplum}{rgb}{0.44, 0.11, 0.11}
\definecolor{smalt(darkpowderblue)}{rgb}{0.0, 0.2, 0.6}
\definecolor{spirodiscoball}{rgb}{0.06, 0.75, 0.99}
\definecolor{turquoise}{rgb}{0.19, 0.84, 0.78}
\definecolor{turquoiseblue}{rgb}{0.0, 1.0, 0.94}
\definecolor{ultramarine}{rgb}{0.07, 0.04, 0.56}
\definecolor{ultramarineblue}{rgb}{0.25, 0.4, 0.96}
\definecolor{violet}{rgb}{0.56, 0.0, 1.0}
\definecolor{violet(ryb)}{rgb}{0.53, 0.0, 0.69}
\definecolor{violet(colorwheel)}{rgb}{0.5, 0.0, 1.0}
\definecolor{zaffre}{rgb}{0.0, 0.08, 0.66}
\definecolor{darkviolet}{rgb}{0.50, 0.0, 0.7}
\definecolor{ltviolet}{rgb}{0.7, 0.0, 0.70}
\begin{document}  
 \title [The quantum superluminality of tunnel-ionization]
 {
 The quantum superluminality in the tunnel-ionization process of H-like atoms 
 }
 \author{Ossama Kullie$^{1}$ and Igor A. Ivanov$^{2}$} 
 \affiliation{
 $^{1}$ Institute for Physics, Department of Mathematics and Natural Science, University of Kassel, Germany}
 \affiliation{$^{2}$ Department of Fundamental and Theoretical Physics, Australian  National University, Australia.}
 \thanks{Email: kullie@uni-kassel.de}
\sffamily 
\begin{abstract} 
 The quantum tunneling time remains the subject of heated debate, 
 and one of its most curious features is faster-than-light or 
 superluminal tunneling.
 Our tunnel-ionization model of the time-delay, presented in previous work, shows good agreement with the attoclock measurement in the adiabatic and nonadiabatic field calibrations, which also enables the determination of the barrier time-delay.  
 In the present work, we show that the tunnel-ionization for H-like atoms with large nuclear charge can be superluminal (quantum superluminality), which in principle can be investigated experimentally using the attoclock scheme.
 We discuss the quantum superluminality in detail for the different regimes of the tunnel-ionization. Our result shows that quantum tunneling faster-than-light is indeed possible, albeit only under somewhat extreme conditions. 
 \end{abstract} 
 \keywords{Ultrafast science, attosecond physics, 
 tunneling and tunnel-ionization time-delay, nonadiabatic effects, 
 weak measurement, interaction time, superluminal tunneling, time and 
 time-operator in quantum mechanics.}  
 \maketitle     
\section{Introduction}\label{sec:int}
\begin{figure}[htp]
\includegraphics[width=20.0cm,height=12.cm]{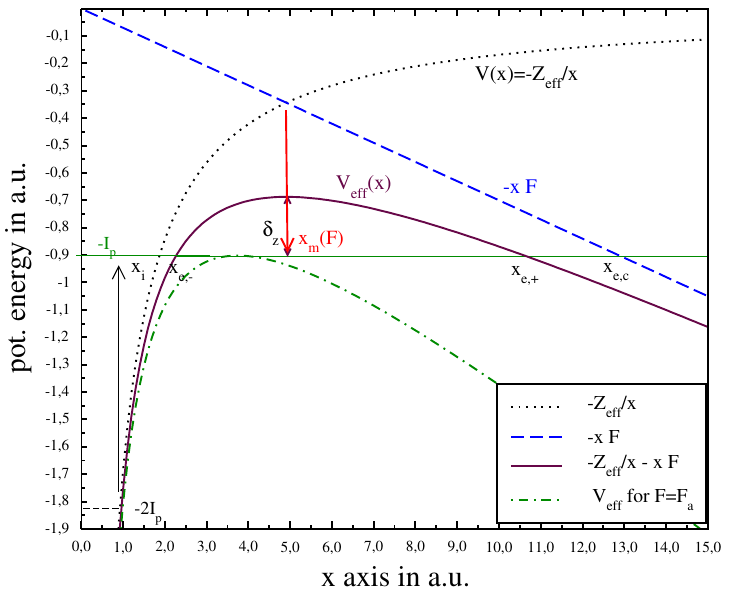}
\centering
\vspace{-6.50cm}
 \caption{\label{figptc}~(Color online) A sketch of the atomic and 
 the effective potential curves, showing the height and width of the 
 barrier formed by the interaction with the laser field.
 The plot is for He-atom in the SAEA model with 
 $Z_{eff}=1.6875$ and $I_{p}\approx0.9\, au$.
 }
\end{figure}

 The interaction of an atom with a laser pulse in strong-field and 
 attosecond physics can be represented in a simplified picture, as 
 shown in Fig. \ref{figptc}, which illustrates the main features of 
 the interaction in the tunnel-ionization regime.
 In previous work we developed in \cite{Kullie:2015,Kullie:2024} 
 a tunneling time-delay model, which explains the measurement result of the 
 attocolck, in the adiabatic \cite{Landsman:2014II} and nonadiabatic 
 \cite{Hofmann:2019} field calibrations.  
 In the model, upon interacting with the electric field of the laser  
 (hereafter field strength) $F$, 
 a direct ionization happen when the field strength reaches a threshold  
 called atomic field strength $F_{a}=I_{p}^{2}/(4 Z_{eff})$ 
 \cite{Augst:1989,Augst:1991}, where $I_{p}$ is the ionization potential  
 of the system (atom or molecule) and $Z_{eff}$ is the effective nuclear 
 charge in the single-active electron approximation (SAEA), 
 and atomic units ($au$) with $\hbar = m = e = 1$ are used throughout the 
 present work.
 However, for $F<F_{a}$ the ionization can happen by tunneling through 
 a potential barrier 
  $V_{eff}(x)=V(x)-x F =-\frac{Z_{eff}}{x}-x F$,
 compare Fig. \ref{figptc}. 
 In the model the tunneling process is described with the help of the 
 quantity $\delta_z=\sqrt{I_{p}^{2}-4 Z_{eff} F}$, where $F$ stands 
 (throughout this work) for {\it the peak electric field strength at 
 maximum}. 
 In Fig.  \ref{figptc} (for details see \cite{Kullie:2015}), the inner 
 (entrance $x_{e,-}$) and outer (exit $x_{e,+}$) points are given by   
 $x_{e,\pm} =(I_{p}\pm\delta_z)/(2F)$, the barrier width is $d_B=
 x_{e,+}-x_{e,-}=\delta_z/F$, and its (maximum) height 
 (at $x_m(F)=\sqrt{Z_{eff}/F}$) is $\delta_z$. 
 At $F=F_{a}$, we have $\delta_z=0$ ($d_B=0$), the barrier 
 disappears and the direct or the barrier-suppression ionization starts, 
 green (dashed-doted) curve in Fig.\ref{figptc}. 
\vskip5pt
\paragraph{Adiabatic tunneling:}\label{pa:adt}
 In the adiabatic field calibration of the attoclock of Landsman et. al. 
 \cite{Landsman:2014II}, we showed in \cite{Kullie:2015} that the 
 tunneling time (T-time)-delay is described by the forms, 
 \begin{eqnarray}\label{Tdi}
 &&\tau_{_{T,d}}=\frac{1}{2(Ip-\delta_z)}, \quad 
 \tau_{_{T,i}}=\frac{1}{2(I_{p}+\delta_z)} 
 \end{eqnarray}
 It was shown that $\tau_{_{T,d}}$ has a good agreement with the 
 experimental result of Landsman et. al. \cite{Landsman:2014II}. 
 $\tau_{_{T,d}}$ is the time-delay of the adiabatic tunnel-ionization
 whereas $\tau_{_{T,i}}$ is the time needed to reach the entrance  
 point $x_{_{e,-}}$. 
 At the limit of atomic field strength,  we have 
 $\lim\limits_{F\to F_{a}} (\delta_z\to 0) \tau_{_{T,d}}=\tau_{_{T,i}}
 = \frac{1}{2Ip}=\tau_{a}$ (see below).  
 For $F> F_{a}$, the barrier-suppression ionization sets up
 \cite{Delone:1998,Kiyan:1991}.
 On the opposite side, we have  $\lim\limits_{F\to 0} (\delta_z\to 
 I_{p}) \tau_{_{T,d}}\to \infty$.  
 So nothing happens and the electron remains in its ground state 
 undisturbed, which shows that our model is consistent. 
 For details, see 
 \cite{Kullie:2015,Kullie:2016,Kullie:2018,Kullie:2020}.
\vskip5pt
\paragraph{Nonadiabatic tunneling:}\label{pa:adtn}
 In the nonadiabatic field calibration of Hofmann et. al. 
 \cite{Hofmann:2019} for He-atom, we showed in \cite{Kullie:2024} that 
 the time-delay is given by   
 \begin{eqnarray}\label{tdion}
 \tau_{dion}(F)
 &=&\frac{1}{2\cdot 4 Z_{eff} }\frac{I_{p}}{F}=\frac{1}{2I_{p}}\frac{F_{a}}{F} 
 \end{eqnarray}
 with good agreement with the experimental result.
 It is easily seen that in the limit $F\to F_{a}, \tau_{dion}=\tau_a$, 
 whereas in the opposite case $F\to 0, \tau_{dion}=\infty$.
 $\tau_a$ is always (adiabatically and nonadiabtically) a lower quantum  
 limit of the tunnel-ionization time-delay and quantum mechanically 
 does not vanish. 
\vskip5pt
\paragraph{Barrier time-delay:}\label{pa:btt}
 Eq. \ref{Tdi} can be decomposed into a twofold time-delay with respect 
 to ionization at $F_{a}$, representing the T-time-delay in an unfolded 
 form, 
\begin{eqnarray}\label{tAd}
\tau_{T,d} 
 &=&\tau_{a}\frac{F_{a}}{F}+ 
 \tau_{a}\frac{F_{a}}{F}\frac{\delta_z}{I_{p}} 
 =\tau_{dion}(F)+\tau_{dB}\equiv \tau_{Ad}
\end{eqnarray}
 From now on, we will refer to it as $\tau_{Ad}\equiv \tau_{T,d}$.
 The first term of Eq. \ref{tAd}, which is identical  
 to Eq. \ref{tdion}, is a time-delay enhancement solely because $F$ is 
 smaller than $F_{a}$, whereas the second term $\tau_{dB}$ is 
 a time-delay enhancement due to the barrier itself, it is the actual 
 T-time-delay, usually referred to as the time spent within the 
 barrier ~\cite{Kullie:2025}, 
 which agrees well with the numerical models of Winful~\cite{Winful:2003} 
 and Lunardi~\cite{Lunardi:2019} and the experimental result based on two field calibrations mentioned above,  as shown 
 in~\cite{Kullie:2025}. 
The conclusion  is that Eq. \ref{tAd} represents a unified T-time picture
(UTTP)~\cite{Kullie:2025,Winful:2003} with $\tau_{dB}= \tau_{dwell}$ and is  determined 
by 
 \begin{equation}\label{tdb}
 \tau_{_{Ad}}-\tau_{dion}=\tau_{dB}=\frac{1}{2}
 \frac{1}{4Z_{eff}} \frac{\delta_z}{F}=\frac{1}{2}
 \frac{d_B}{4Z_{eff}}
 \end{equation} 
 Obviously, Eq. \ref{tdb} gives 
 the time spent within the barrier $\tau_{barrier}$~\cite{Kullie:2025,Winful:2003,Lunardi:2019} 
 or the dwell time $\tau_{dwell}$~\cite{Winful:2003}.
 It shows a linear dependence of the $\tau_{dB}$ on the 
 barrier width $d_{B}$, which tends to zero in the limit of $F=F_{a} 
 (\delta_z=0)$ because the barrier disappears~\cite{Hartman:1962}.
 A similar linear dependence on the barrier width is shown by Balcou~\cite{Balcou:1997} for the phase-time in optical tunneling. 
 Unfortunately, $\tau_{dB}$ cannot be measured by the experiment directly 
 since the first term $\tau_{dion}$ is always present. 
 However, as seen in Eq. \ref{tdb}, it can be determined taking into 
 account both field calibrations, as shown in~\cite{Kullie:2025}.
 See further below Fig. \ref{figQSbtd}.
\vskip5pt
\paragraph{thick-barrier and Weak measurement}\label{ssec:wm}
 In the limit of a thick (or opaque) barrier, the barrier width approaches the so-called 
 classical barrier width $d_C$,  $\lim\limits_{F\ll F_{a}} 
 d_B \approx I_{p}/F=d_C$ ($x_{e,c}$), compare Fig. \ref{figptc}. 
 In this limit the barrier height 
 $\lim\limits_{F\ll F_{a}} {\delta_z\to I_{p}}$, and the barrier time 
 \begin{eqnarray}\label{tdbl}
 \lim\limits_{F\ll F_{a}}\tau_{dB}&=&\tau_{a}
 \frac{F_{a}}{F} \frac{(\delta_z\approx I_{p})}{I_{p}}
 \approx  \tau_{a} \frac{F_{a}}{F}=\tau_{dion}\\\nonumber
 &=&\frac{1}{2}
 \frac{1}{4Z_{eff}} \frac{I_{p}}{F}=\frac{1}{2}
 \frac{d_C}{4Z_{eff}}
 \end{eqnarray}
 The result shows that the barrier time-delay 
 $\tau_{dB}$ for a thick-barrier (or opaque barrier) is approximately 
 equal to tunnel-ionization time-delay $\tau_{dion}$ of Eq. \ref{tdion}. 
 In addition, the back reaction of the measurement of the system can 
 be found from $\tau_{T,i}$ as the following
 $ \lim\limits_{F\ll F_{a}} \tau_{T,i}=
  \frac{1}{2}\frac{(I_{p}-\delta_{z})}{4Z_{eff}F} 
 =\frac{1}{2}\frac{\varepsilon_{_{F}}}{4Z_{eff}F}=\tau_{backr}, 
 $
 where $\varepsilon_{F}= \delta_{z}-I_{p}$ is small under the condition 
 a weak measurement ($F\ll F_{a}$). 
 However, the condition of WM is not necessary and we can assume that  
 $\tau_{backr}$ always represents the back reaction of the system,
 although it can be generally interpreted as the time needed to reach the 
 barrier entrance in the strong-field interaction as already 
 mentioned~\cite{Kullie:2025}. 
\section{The Quantum Superluminaliy}\label{sec:qs}
 Superluminal tunneling or (quantum) superluminality, which means that the 
 T-time of a barrier can be a shorter than that of light passing 
 through a vacuum over the same distance of the barrier width 
 (superluminal speed), is one of the most exciting and controversial 
 phenomena in quantum physics. 
 In photonic tunneling many authors reported that they measured 
 superluminal T-time experimentally.
 They also argued that it does not violate the principle of causality  
 or special relativity  
 \cite{Heitmann:1994,Nimtz:2008,Dumont:2023,Chiao:1997,Chiao:1999}.
 In his explanation of tunneling of light pulses through a barrier    
 (Nature of ``Superluminal" Barrier Tunneling)
 \cite{Winful:2003,Winful:20031,Winful:20032},   
 Winful argued that the measured T-time is not a propagation time and, 
 if interpreted correctly, no superluminal transport would be observed.
 He claims that what is really observed then is a phase shift in an 
 amplitude modulation.
 B\"uttiker et. al. \cite{Buettiker:2003} believe that it is just an 
 artifact of our definition of speed as marked by the arrival of the 
 peak in the pulse, whereas Steinberg et. al. \cite{Steinberg:1993} 
 suggest that the group delay (or "phase time") gives a better 
 description of the physically observable delay.

 In quantum tunneling of a stream of particles that scatters at
 a potential barrier, Winful \cite{Winful:2003} (see sec. \ref{pa:btt}) 
 showed that the group delay is equal to the dwell time plus 
 a self-interference time-delay.  
 The Hartman effect in quantum tunneling, which says that the phase 
 time of tunneled particles is independent of barrier width for 
 sufficiently wide (thick) barrier, is explained on the basis of  
 saturation of the integrated probability density (or number of 
 particles) under the barrier. 
 More recently Dumont et. al. \cite{Dumont:2023} investigate the 
 MacColl-Hartman (or Hartman) effect \cite{MacColl:1932,Hartman:1962}. 
 They showed that the phase time is the appropriate way of measuring 
 the barrier interaction time and it can lead to faster-than-light 
 time.
 However, they claim that only the statistics of the arrival time of 
 the particles distribution can be used to make statements.
 They conclude that their analysis leads to the impossibility of 
 superluminal signaling using the MacColl-Hartman effect~\cite{Dumont:2023}.
 
 Nanni~\cite{Nanni:2023}, on the other hand, recently reported the 
 possibility of superluminal transport of electromagnetic energy in 
 photonic tunneling.
 And what's even more exciting: Lunardi et. al.~\cite{Lunardi:2019} provide, a plausible explanation for a possible superluminal tunneling process in the attoclock framework using numerical simulation. 
 
 Based on our model, below we study the superluminality in the attoclock 
 framework, where we refer to it as quantum superluminality (QS), with 
 the conclusion that superluminality is quite possible, albeit under 
 somewhat “extreme” conditions.
 The advantage of our model in explaining the attoclock measurement 
 lies in the simplicity of determining the tunnel-ionization time-delay  
 in the adiabatic and nonadiabatic field calibrations and allows the  
 extraction of the barrier time-delay.
 It gives a clear picture of tunneling in general and emphasizes the 
 UTTP in particular, in line with the approaches of Winful~\cite{Winful:2003} 
 and Lunradi~\cite{Lunardi:2018,Lunardi:2019} explained above, 
 see sec. \ref{pa:btt}.
 \begin{figure}[t]
 \vspace{-1.50cm}
 \hspace{-1.0cm}
\includegraphics[scale=0.45]{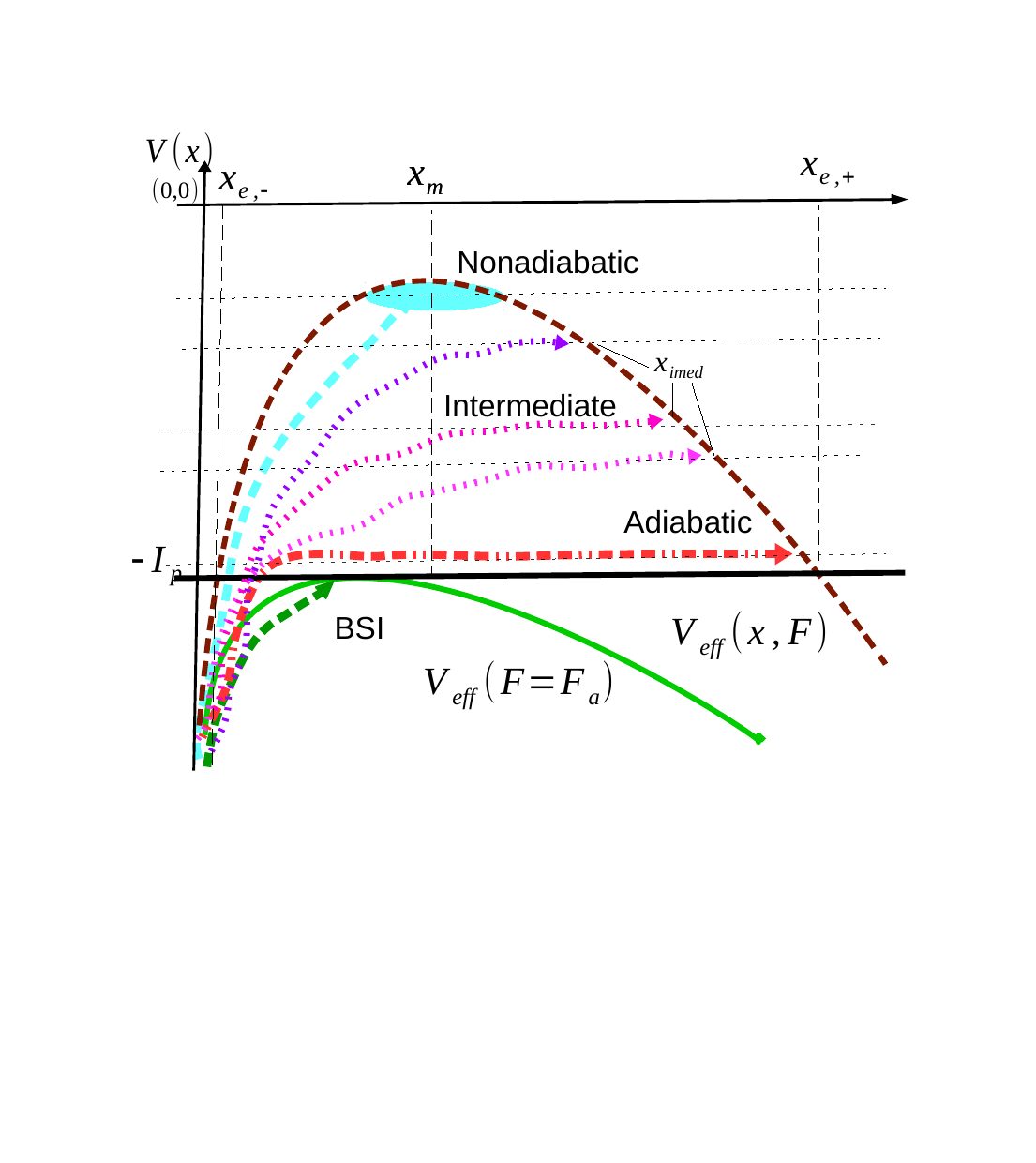}
 \vspace{-3.20cm}
 \caption{\label{figTI} (Color online) 
 Outline of tunnel-ionization in three cases: adiabatic (dashed-dotted, 
 red), intermediate (dotted, magenta, purple) and nonadiabatic (dashed, 
 light blue). 
 As well as the barrier-suppression ionization at $F_{a}$ (green).} 
\end{figure}
\vskip5pt
 The measurement of the attoclock experiment offers two field 
 calibrations based on adiabatic or nonadiabatic tunneling theory.
 However, the tunneling process can also be described by an 
 intermediate regime in which adiabatic and nonadiabatic 
 tunnel-ionization coexists \cite{Ivanov:2005}. 
 This is shown in Fig. \ref{figTI} by dotted (magenta or purple 
 colored) curves.
 To investigate the QS, we first discuss the adiabatic and 
 nonadiabatic cases for which experimental results are available
  and later will deal with the intermediate regime,  which is formulated into a generalized model using a switching parameter. The final result will then be presented based on this model.
 \subsection{QS in adiabatic tunnel-ionization}\label{pa:ad}
 \paragraph{Barrier time-delay} Adiabatic tunneling happens along the so-called horizontal channel, as 
 shown in Fig. \ref{figTI} (dashed-dotted red colored), where the exit 
 point is at $x_{e,+}$ (compare Fig. \ref{figptc}) and the barrier 
 width is $d_{B}=\frac{\delta_z}{F}$.
 The barrier time-delay itself $\tau_{dB}=\frac{d_B}{8 Z_{eff}}$ 
 shows a linear dependence on the barrier 
 width \cite{Hartman:1962,Kullie:2025}, in contrast to the Hartman 
 effect, i.e. the lack of dependence of T-time-delay on barrier 
 length for thick (opaque) barrier as discussed in \cite{Winful:20031}.
 Therefore, it is to be expected that, within the framework of the 
 attoclock, even in the limit of a thick-barrier, no 
 superluminal tunneling takes place.
 This is indeed true in experiments when considering the light atoms 
 under investigation, such as the He atom as shwon in
 ~\cite{Landsman:2014II,Landsman:2015,Kullie:2015}.  
 It can be seen if one compares the time-delay $\tau_{dB}$ with the 
 time required for the light, to overcome the barrier width $d_B$ with 
 the speed $c$ in vacuum $\tau^{Ad}_{c}= \frac{d_B}{c}$, where one can 
 easily verfy that $\tau_{dB}/\tau^{Ad}_{c}=\nicefrac{c}{(8 Z_{eff})}>1$ 
 for $Z_{eff}=1.6875$~\cite{Clementi:1963} (or $Z_{eff}=\sqrt{2 I_{p}}=1.344$). 
 \begin{figure}[t]
  \vspace{-1.0cm}
  \centering
  \includegraphics[scale=0.4]{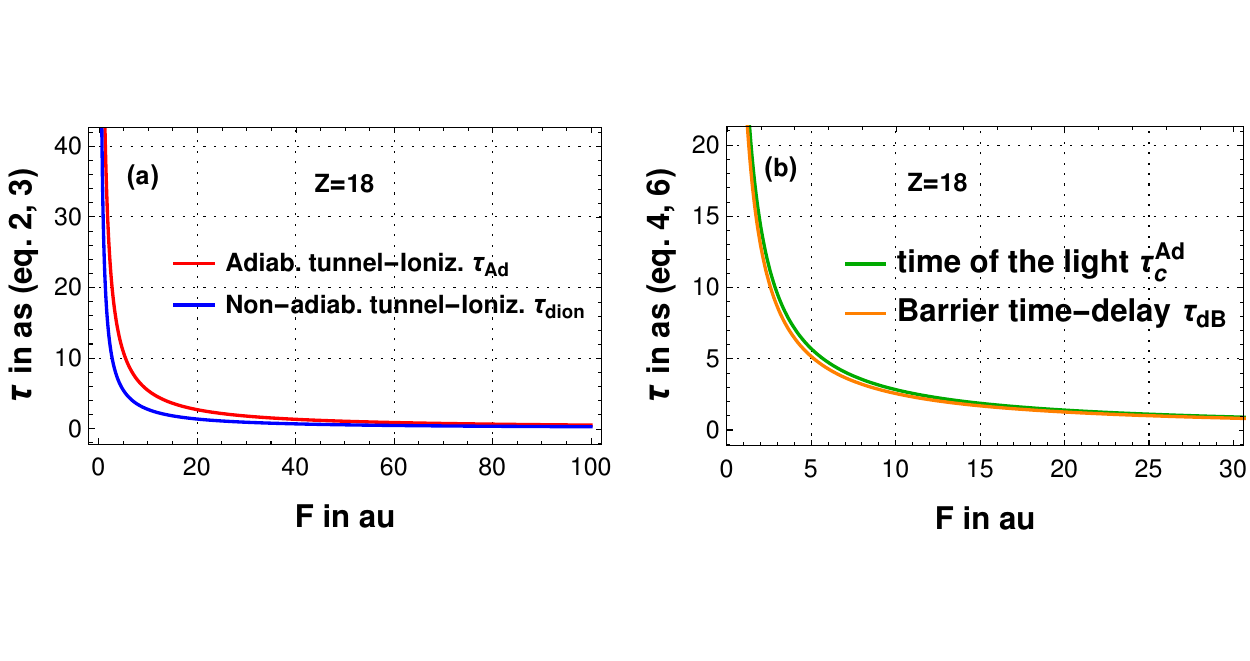}
    \vskip-30pt
  \caption{\label{figQSbtd} (Color online)
 (a): Tunnel-ionization time-delay in the adiabatic and  nonadiabtic field calibration $\tau_{Ad}, \tau_{dion}$ vs. field strength $F$ (in au) for $Z=18$. 
 (b): Time-delay of the barrier $\tau_{dB}$ (green)  and the time of light $\tau_{c}^{Ad}$ (orange) vs. field strength $F$ (in au) for $Z=18$, see Eq. \ref{zvalue}. 
 Hereafter "as" is the abbreviation for attosecond.}
 \end{figure}

 However, a closer look at $\tau_{dB}$ and $\tau^{Ad}_{c}$ indicates 
 that the tunneled electron traverses the barrier region with an 
 average speed of $v_{av}=8\,Z_{eff}$.
 The comparison of $\tau_{dB}$ (Eqs.~\ref{tdb}-\ref{tdbl}) and the corresponding travel time of the light $\tau^{Ad}_{c}$ leads to the following condition for the occurrence of superluminal tunnel-ionization, 
 \begin{equation}\label{zvalue}
  Q_{dB}=\frac{\tau_{dB}}{\tau^{Ad}_{c}}=\frac{c}{8 Z_{eff}}< 1\Rightarrow 
  Z_{eff}> {c}/{8}\approx 17.13
 \end{equation} 
 where $c=173.0359992$ (CODATA22). 
 This indicates that the barrier tunneling time-delay $\tau_{dB}$ can be superluminal,  or that the QS occurs when the strength of the nuclear potential large and $Z_{eff}\ge 18$ holds. 
 Fig. \ref{figQSbtd} (a) shows $\tau_{dion}$ (Eq.~\ref{tdion}) and $\tau_{Ad}$ (Eq. \ref{tAd}) as a function of the field strength for one-electron H-like ionic Argon ${\rm Ar}^{+17}$ where $Z_{eff}=Z=18$, while their difference, i.e. the barrier time-delay $\tau_{dB}$ (Eq. \ref{tdb}), is shown in Fig.~\ref{figQSbtd} (b) together with the light travel time $\tau_{c}^{Ad}$ through the barrier width (see text above Eq. \ref{zvalue}), from which it is clearly evident that the barrier time-delay $\tau_{dB}$ becomes smaller than $\tau_{c}^{Ad}$ for $Z\ge 18$. Indeed, $\tau_{dion}$ and $\tau_{Ad}$ become very small for large field strength $F$, but Eqs. \ref{tdb}-\ref{tdbl} and the corresponding condition of Eq. \ref{zvalue} are always satisfied for $F\le F_{a}$, see further below.  In \ref{sec:apdx2} we present result from the numerical integration of the time-dependent Schr\"odinger equation (NITDSE). For $Z=18$ and $F=50/\sqrt{2}\, a.u., \omega=3\, a.u.$, we obtained a time-delay of $\tau\approx 4 as$. Unfortunately, this differs from the result obtained from our model $\tau_{dB}\approx \tau_{dion}\approx 0.7 \, as$. we are not able to explain this discrepancy at the moment. 
 
 For the rest of the present work, we discuss H-like atoms 
 ${\rm A}^{+(Z-1)}$ in the ground state ($Z_{eff}=Z$), where $Z$ is 
 the nuclear charge of the atom $\rm A$, e.g. ${\rm He}^{+},\, 
 {\rm Li}^{+2}, ...$, with the nonrelativistic ionization potential 
 $I_p=\nicefrac{Z^{2}}{2}$.
 In this case, the barrier height 
 $\delta_z=\sqrt{I_{p}^{2}-4 Z\,F}=\nicefrac{Z^{2}}{2}
  \sqrt{1-\nicefrac{F}{F_{a}}}
 =\nicefrac{Z^{2}}{2}\sqrt{1-\nicefrac{16 F}{Z^3}}$
 increases with $Z^2$ 
 (relativistic effect are not crucial and can be taken into account, 
 see below).
 For large $Z$ the second term under the square root is small for 
 $F\ll F_{a}=\nicefrac{Z^3}{16}$ and the barrier width $d_B=
 \nicefrac{\delta_z}{F}$ increase with $Z^2$ for fixed $F$ and both the  
 height and width of the barrier become too large for large $Z$.
 In general, for large $Z$ in the Attoclock framework, where 
 $F\ll F_{a}=\nicefrac{Z^3}{16}$, the limit of a thick (or opaque) 
 barrier applies.

\begin{figure}[t]
  \centering
  \includegraphics[scale=0.40]{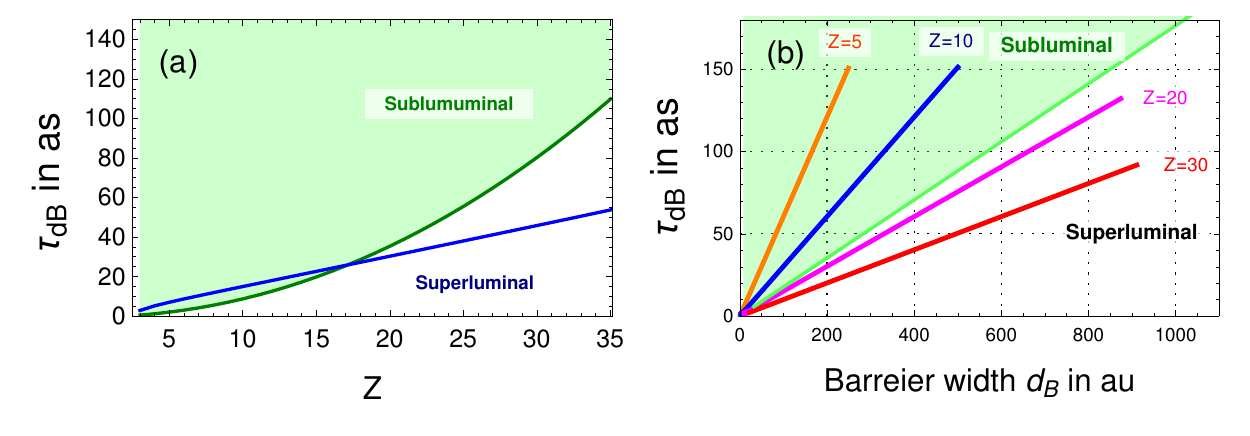}
  \vskip-15pt
  \caption{\label{figQSbtd1} (Color online)
 (a): tunnel-ionization  $\tau_{dB}$ (Eq. \ref{tdb}) vs. $Z$ for $F=1 au$, 
 The green/white area is subluminal/superluminal. 
 (b): $\tau_{dB}$ vs. barrier width $d_{B}(F)$ for different 
 $Z$-values. The boundary between subluminal area (green) and  
 superluminal area (white) is along $Z=c/8$, see 
 Eq. \ref{zvalue}.}
 \end{figure}

 In Fig. \ref{figQSbtd1}~(a), we let $Z$ increases as a parameter and plot 
 $\tau_{dB}(Z)$ (blue) together with $\tau_{c}^{Ad}(Z)$ (green), where  
 we used the value of $F=1\, au$ . 
 It  is clearly seen that the barrier time-delay is subluminal for $Z< 18$ (green area). 
 Whereas the blue colored line or $\tau_{dB}$ becomes superluminal for $Z\ge 18$ (white area). The part of the blue curve in the green area applies for $Z\le 17$ and the 
 intersection point marks the value $Z = \nicefrac{c}{8}$.
 It is worth noting that a change in $F$-value does not alter the 
 overall appearance in Fig.~\ref{figQSbtd1}~(a) provided we choose $F< F_a$  
 for the smallest $Z$ in each case, where we started in  Fig.~\ref{figQSbtd1}~(a)
 with $Z=3$ ($F_a\approx 1.7\, au$). 
 
 In Fig. \ref{figQSbtd1}~(b), we plot $\tau_{dB}$ vs. the barrier 
 width 
 for various $Z$-values.
 As can be seen in Fig. \ref{figQSbtd1}~(b), $\tau_{dB}$ increases linearly 
 with the barrier width $d_B$. 
 The green colored area represents the subluminal tunnel-ionization, 
 and the boundary $\tau^{Ad}_{c}$ with $Z=c/8$ (see Eq.~\ref{zvalue}) 
 marking the transition to superluminal tunnel-ionization. 
 For $F\rightarrow F_{a}$, we have $\tau_{dB}=\tau^{Ad}_{c}=0$ 
 because the barrier disappears, as is known from the Hartman effect. 
 However, the linear dependence of $\tau_{dB}$ on the barrier width 
 contrasts the Hartman-effect for thick-barrier (with $F\ll 
 F_a\approx Z^{3}/16$), which makes the above mentioned argument of Winful~\cite{Winful:20031} not applicable (first paragraph of sec.~\ref{sec:qs}).
 On the other hand, for a fixed barrier width $d_{B}$, $\tau_{dB}$ 
 decreases with increasing barrier height $\delta_z$ (increasing $Z$), 
 as can be seen form  Eq.~\ref{tdb} (a vertical cut line for a chosen 
 $d_B$-value in Fig.~\ref{figQSbtd1}~(b)).
 This agrees with the observation of Spierings et.~al.~\cite{Spierings:2021}.
 Using the Larmor clock (LC) measurement for atoms tunneling through 
 an optical barrier, they found that atoms generally spend less time tunneling through a higher barrier. 
 It appears that we are dealing here with a universal behavior that is consistent with the time-energy uncertainty principle.
 
 \paragraph{Adiabatic time-delay}\label{pa:adb}
 Our comparison so far concern only the barrier time-delay, while the measurement in the adiabatic field calibration provides the time-delay $\tau_{Ad}$ given in Eq.~\ref{tAd}, which is the relevant (measurement) physical quantity to be compared with $\tau^{Ad}_{c}$. It includes the time-delay $\tau_{dion}$ and the barrier time-delay $\tau_{dB}$ given in Eq.~\ref{tAd}. 
 This shifts the boundary of QS of the adiabatic time-delay of tunnel-ionization that can be experimentally measured. For thick-barrier $\delta_{z} \approx I_{p} \Rightarrow\tau_{Ad}\approx \frac{I_{p}}{4 Z\, F}$, 
 and after some manipulation, as previously done, we obtain that QS occurs when
 \begin{equation}\label{qsad}
  Q_{ad}=\tau_{Ad}/\tau_{c}^{Ad}<1  \Rightarrow Z> c/4
  \approx 34.26
 \end{equation}
 This shows that quantum superluminality is possible in adiabatic 
 tunneling, albeit under somehow extreme conditions $Z\ge 35$ where the 
 probability of the tunneling in question is most likely very low, as 
 is known from quantum tunneling~\cite{Dumont:2023,Winful:2006II,Lunardi:2019}.  
 
 We conclude that QS is a fundamental property of adiabatic tunnel-ionization, since our model is consistent and exhibits a universal behavior in accordance with the unified tunneling picture proposed by Winful~\cite{Winful:2003,Lunardi:2018,Lunardi:2019},  and is well supported by the agreement of the barrier time-delay with  the LC time-delay measurement~\cite{Spierings:2021}, compare  sec.~\ref{ssec:wm} the limit of thick-barrier.
 \subsection{QS in nonadiabatic tunnel-ionization }\label{pa:nad}
 Nonadiabatic tunnel-ionization happens along the vertical channel, as 
 shown in Fig.~\ref{figTI} (light blue curve), where the exit point is 
 $x_{m}=\sqrt{\nicefrac{Z}{F}}$ at the top of the barrier.
 This changes the situation as the width traversed becomes significantly 
 shorter $d_{{m}}\approx x_{m}$, where the initial point $x_{e,-}$ is 
 small and can be neglected. 
 In this case the time-delay is given by $\tau_{dion}=\frac{1}{8 Z}
 \frac{I_p}{F}\approx\frac{Z}{16 F}=\frac{1}{16} x^{2}_{m}$ 
 (the relativistic effect changes it slightly).  
 We will first compare our $\tau_{dion}$ with $\tau^{Ad}_{c}=d_B/c$ and later do the same  for $\tau^{Nad}_{c}=\frac{d_{{m}}}{c}\approx\frac{x_{m}}{c}$,
 where the traversed width of the non-adiabatic tunnel-ionization (the 
 vertical channel) at the exit point $d_{{m}}=x_{m}$ is assumed.
 \vspace{0.2cm}
 \paragraph{With $\tau^{Ad}_{c}$:}\label{pa:nada}
 For thick-barrier $\delta_z\approx I_{p}, d_{B}\approx d_{C}
 =\nicefrac{I_{p}}{F}, \tau^{Ad}_{c}={d_C}/{c}$ and after some manipulation, we get a similar result to Eq. \ref{zvalue} ($Z<17.13$) showing that QS is possible 
 in the nonadiabatic tunnel-ionization. 
 However, $\tau_{dion}$ in the nonadiabatic tunnel-ionization is an  
 interaction time (multiphoton absorption) and not a transverse time 
 along  the barrier width $d_{B}$ ($x$-axis direction). 
 The comparison of $\tau_{dion}$ (vertical channel) with the travel 
 time of the light through the barrier width $d_{B}$ (horizontal 
 channel) is fictitious in this case; in other words, our comparison 
 with $\tau^{Ad}_{c}$ is inappropriate.
 
 
 
 \vspace{0.2cm}
 \paragraph{With $\tau^{Nad}_{c}$:}\label{pa:nadb} 
 Rather, one has to compare $\tau_{dion}$ with $\tau^{Nad}_{c}=d_{{m}}/c
 \approx x_{m}/c$, which is the time it takes for the light to 
 traverse the distance $d_{{m}} \approx x_{m}$ horizontally. 
 In this case  with $I_{p}\approx Z^2/2$, QS is possible if  
 \begin{eqnarray}\label{qsnad}
Q_{Nad}&=&\tau_{dion}/\tau^{Nad}_{c}=\frac{I_{p}}{8 Z\, F}\frac{c}{x_{m}}\\\nonumber
 &=&\frac{I_{p}\, c}{8 Z\, F} \sqrt{F/Z}
 \approx\frac{c}{16} \sqrt{\nicefrac{Z}{F}}=\frac{c}{16} x_{m}<1
 \end{eqnarray}
 This leads to the condition $F> (c/16)^2\cdot Z\approx 73.36\,Z$, 
 and together with the condition $F\le F_{a}\approx \frac{Z^{3}}{16}$ (the tunneling regime) we get, 
\begin{eqnarray}\label{Fnonad}
F_{c}&=&({c/16})^{2} Z< F\le{Z^{3}}/{16}\\\nonumber 
&& \Rightarrow ({c/16})^{2} Z<{Z^{3}}/{16} \Rightarrow Z>34.26
\end{eqnarray}
(the relativistic effects change it slightly.) 
 For such large $Z\ge 35$, we have $F\ge 35\, (c/16)^{2}$, and the 
 corresponding intensities are on the order of $ 9.0\, 10^{19}
 (W/cm^{2})$,  meaning that a laser pulse with very high intensity 
 is required. 
 The second condition of $F<F_a\approx Z^{3}/16$ for $Z\ge 35$ corresponds to 
 $F_{a}= 9.4*10^{19} (W/cm^{2})$ (with relativistic effect  $F_{a}=9.7*10^{19} (W/cm^{2})$). 
 However, the thick barrier limit does not need to be taken into account, as  this is a multiphoton process (vertical channel).   
 \begin{figure}[t]
 \hspace{-0.7cm}
\includegraphics[width=8.0cm]{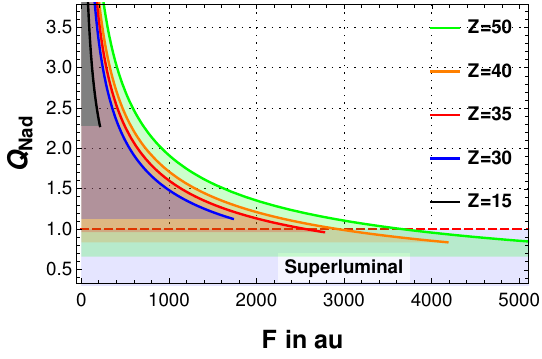}
 \vspace{-0.30cm}
 \caption{\label{figQnadb} (Color online) 
 $Q_{Nad}$ (see Eq. \ref{qsnad}) vs. field strength $F$ (in au) for various $Z$ values. 
 In the colored area below the red dashed line, tunnel-ionization is superluminal, which is the case when $Z \ge 35$.}
\end{figure}
  In Fig. \ref{figQnadb}, we plot $Q_{Nad}$ of Eq. \ref{qsnad} vs. field strength $F$ for various $Z$.  
  For every $Z$ we let $F$ vary up to  $F_{a}(Z)$, where $F_{a}(Z)=I_{p}^{2}/(4 Z)$. 
 However, as given in Eqs.~\ref{qsnad},~\ref{Fnonad} only when $F_{c}=(c/16)^2\,Z< F \le F_{a}$ is satisfied supeluminlaity exists, which is conform with the condition $Z\ge 35$ as given in Eq.~\ref{Fnonad} and clearly seen in fig. \ref{figQnadb} (red, orange and green colored curve).  For $Z<35$ we have $F_c>F_{a}$ and no superluminlaity can exist (e.g. for $Z=15, 30$ in fig. \ref{figQnadb}, the black and blue colored curves). 
 For $F>F_{a}$ the barrier-suppression-Ionization (BSI) region is reached, which will be the subject of future work. 
 
 
 This again shows that the QS of tunnel-ionization requires extreme 
 conditions, now when comparing the time delay $\tau_{dion}$ with 
 $\tau^{Nad}_{c}$, the time it takes for light to 
 travel (horizontally) the same distance $x_{m}$ that the particle 
 travels horizontally while climbing the potential barrier (vertical 
 channel), compare Fig.~\ref{figTI}. 
 Thus, the time-delayed tunnel-ionization of the vertical channel (or 
 the multiphoton process), which is an interaction time-delay (as 
 already mentioned), can be superluminal in the attoclock framework, 
 albeit only under extreme condition, as we have seen in the adiabatic 
 tunneling in sec.\ref{pa:adb}. 

 
 
 Concerning the interaction time in this case, the tunnel-ionization 
 (vertical channel) occurs by multiphoton absorption of a number of 
 photons $n$, and  we can assume that $n\approx I_{p}/\omega$. 
 The time-delay can be written in the form $\tau_{dion}=\frac{1}{8 Z}
 \frac{n\omega}{F}$, meaning that for the absorption of $n$ photons 
 within the atomic potential we have
 \begin{equation}\label{tph}
 \tau_{nph}= n \frac{\omega}{8 Z\,F}=
 n \frac{\pi}{4 Z}\frac{1}{F T}=
 n\frac{1}{\alpha}\frac{\pi}{4 Z}\frac{1}{\lambda\, F}, 
 \end{equation}
which can also be written in the form $\tau_{dion}=n\, \tau_{1ph}$, where $\tau_{1ph}$ is the time for the absorption of one photon, where $\alpha$ is the fine structure  constant ($\alpha$ is the strength of the light-matter interaction), $T$ the period and $\lambda$ the wave length of the laser pulse. 
 The relation \ref{tph} shows that the time-delay $\tau_{dion}$ 
 decreases with increasing strength of the potential, $Z$, and the 
 period of the laser pulse $T$. 
 The later property is in contrast to the finding of Esposito 
 \cite{Esposito:2001} and Nimtz \cite{Nimtz:2008} that the time-delay 
 (phase time) for opaque barrier correlate with reciprocal of the 
 photonic frequency ($\tau\sim 1/\nu=T$) in photonic tunneling.
 Furthermore, $\tau_{dion}$ depends linearly on the barrier width (i.e. 
 on $\frac{I_{p}}{F}=d_{C} \approx d_{B}$ for thick-barrier, compare 
 Eq. \ref{tdion}), which is why Winful's argument mentioned above at the 
 beginning of this section (based on the lack of dependence on the 
 barrier width \cite{Winful:20031}) is not applicable. 
 \subsection{The intermediate regime}\label{pa:in} 
 This is an interesting regime because in this case the two 
 tunnel-ionization channels co-exit \cite{Ivanov:2005}, adiabatic 
 (horizontal channel) and nonadiabatic (vertical channel), as depicted 
 in Fig. \ref{figTI} dotted (magenta, purple) curves. 
 Interesting because, as we found in sec. \ref{pa:ad}, QS is possible 
 in the adiabatic case, but the probability of transmitted particles 
 can be very low.
 While, as we found in sec \ref{pa:nad}, QS is hardly possible in 
 nonadiabatic tunnel-ionization, where the probability of transmitted 
 particles can be assumed to be high (multiphoton absorption), compare 
 Fig. \ref{figTI} dashed (light blue) curve.
 Therefor, the intermediate tunnel-ionization can vary with different 
 contributions between the two channels, opening the possibility of 
 QS with different probabilities of the transmitted particles.
 Nevertheless, we can reasonably describe this regime based on our model, which accurately describes the two regimes mentioned above and our understanding of the last paragraphs \ref{pa:ad}, \ref{pa:nad}.
 It is clear from our discussion that the first term in the model 
 \ref{tAd} ($\tau_{dion}$)
 always exist, compare sec \ref{pa:btt}, \ref{pa:ad} and \ref{pa:nad}.
 The second term $\tau_{dB}$, the barrier time-delay, which exists in   
 the adiabatic case (Eq. \ref{tAd}), is reduced in the intermediate 
 regime so that it disappears by reaching the vertical channel ot the 
 nonadiabatic case, compare Fig. \ref{figTI}.
 \subsubsection{Establishing a model for the intermediate regime}\label{sec:min}
 Before we discuss the QS in the intermediate regime in the next subsection, we first introduce the model for studying QS in the intermediate regime in this subsection.
 
 In the intermediate regime, tunnel-ionization can be divided into two 
 parts: partially climbing the barrier to a virtual state 
 corresponds to a contribution from the vertical channel, and crossing 
 the remaining barrier corresponds to a contribution of the horizontal 
 channel, reaching the tunnel exit point ($x_{imed}$), as shown in 
 Fig. \ref{figTI}, where three such events are shown (dotted purple or 
 magenta curves). The situation can be described by a dimensionless switching parameter $0\le\zeta\le 1$ and with Eq. \ref{tAd} one obtains,
 \begin{eqnarray}\label{timed}
  \tau_{imed}&=&\tau_{dion} +\zeta \tau_{dB}
  = \frac{I_{p}}{8 Z_{eff}F} 
 +\zeta \frac{\delta_z}{8 Z_{eff}F},\\\label{timed1}
 d_{imed}&=&x_{m}+\zeta (d_{B}-x_{m})=(1-\zeta)  
 x_{m}+\zeta\,d_{B}, 
 \end{eqnarray}
 where $Z_{eff}=Z$, $d_{B}$ is the barrier width, $\tau_{dB}$ is the barrier time-delay of Eq. \ref{tdb}, and  $\zeta=0,1$ correspond to the nonadiabatic and 
 adiabatic case, respectively. 
 The barrier width in this case, denoted $d_{imed}$ with the exit point 
 $x_{imed}$ (compare Fig.~\ref{figTI}), is approximated using the 
 same parameter $\zeta$, which is justified because the tunnel-ionization 
 time-delay (in particular $\tau_{dB}$, Eq.~\ref{tAd} and~\ref{tdb}) 
 depends linearly on the barrier width.
 This is justified as long as $F\le F_{a}$ holds, where $d_{B}\ge x_{m}$; furthermore, our focus is on the limit $F\ll F_{a}$, compare fig \ref{figQSbtdzeta}. 
 The limits $\zeta=0$ and $\zeta=1$ correspond to the two cases given 
 in~\ref{pa:nadb},~\ref{pa:ad}, since $ \lim\limits_{\zeta\to 0,1} 
 \tau_{imed}=\tau_{dion}, \, \tau_{Ad}$, and $\lim\limits_{\zeta\to 0,1} 
 d_{imed}=x_{m},\, d_{B}$ respectively. 

 To verify the reliability of our model in  Eq.~\ref{timed1}, we plot in Fig.~\ref{figQSbtdzeta}~(a) the exit point  $x_{exit}$ of different approaches for hydrogen atom versus field strength $F$. We take the limit of thick barrier, where $d_B=d_C$ (which is commonly used in the literature) and $d_{imed}\approx x_{e,imed}$. As a test we set $\zeta=0.5$ and plot $x_{e,imed}(\zeta=0.5)$ (denoted PW with $\zeta=0.5$, dark green colored in Fig.\ref{figQSbtdzeta}~(a)) together with a model of exit point of NITDSE recently published in \cite{IAIvanov:2025}  by Ivanov et. al. (red colored with error bars, see ref. \cite{IAIvanov:2025}), and a model of Li et al \cite{Li:2016} (black colored in Fig.\ref{figQSbtdzeta}~(a)).  
 Also shown the exit points of the adiabatic $x_{e,+}$ (light green) and nonadiabatic $x_m$ (orange) cases of our model (see sec. \ref{sec:int}). 
 The agreement of our model for the hydrogen atom with $\zeta=0.5$ as a test with the results given in the literature, in particular the NITDSE from ref. \cite{IAIvanov:2025}, shows that our model is reliable and can serve as a generalized model, although $\zeta$ is a parameter that needs to be precisely determined. Further information, possibly from experiments, is required for this.
\vskip5pt
\subsubsection{QS in the intermediate regime}\label{sec:stin}
We will investigate the QS for the intermediate tunnel-ionization with the 
 help of Eqs.~\ref{timed},~\ref{timed1} which, as we will see, shows 
 interesting features.
 Its behavior at the boundaries $\zeta=0,1 $ coincides with the two 
 cases given of sec.~\ref{pa:nadb}, \ref{pa:adb}, respectively. 
 As before, we take the thick-barrier  ($F\ll F_{a}$), where $\delta_{z}\approx I_p, \, 
 d_{B}\approx d_{C}= {I_{p}}/{F}$, and first compare $\tau_{imed}$ 
 with $\tau_{c}^{Ad}=d_{C}/c=I_{p}/cF$, later with $\tau_{c}^{imed}=d_{imed}/c$.
\vskip10pt
 \paragraph{With $\tau_{c}^{Ad}$:}\label{pa:ina}
 In this case and from Eq. \ref{timed}, we obtain the following condition for superluminality,  
 \begin{equation}\label{qsimeda}
 Q_{imed}^{a} =\frac{\tau_{imed}}{\tau_{c}^{Ad}} 
  =\frac{c(1+\zeta)}{8Z}\le 1 ,\Rightarrow Z\ge\frac{c\, (1+\zeta)}{8}  
 \end{equation}
 which leads to $Z\ge Z_{\zeta}={(c/8)\, (1+\zeta)}$, where 
 $Z_{\zeta}\approx 17.13-34.26$ for $0\le \zeta\le 1$ and field strength 
 $F< F_{a}$ (or $F\ll F_{a}$ thick-barrier).
 The tunnel-ionization is for $Z\le 17$ subluminal, see sec. \ref{pa:nada}.
 
 An increase in $\zeta$ increases the $Z$ value (equation \ref{qsimeda}) and the adiabatic contribution, with the occurrence of superluminal speed depending on the $\zeta$ value.
 \begin{figure}[htp]
 \vskip-25pt
 \includegraphics[height=5.cm,width=8.50cm]{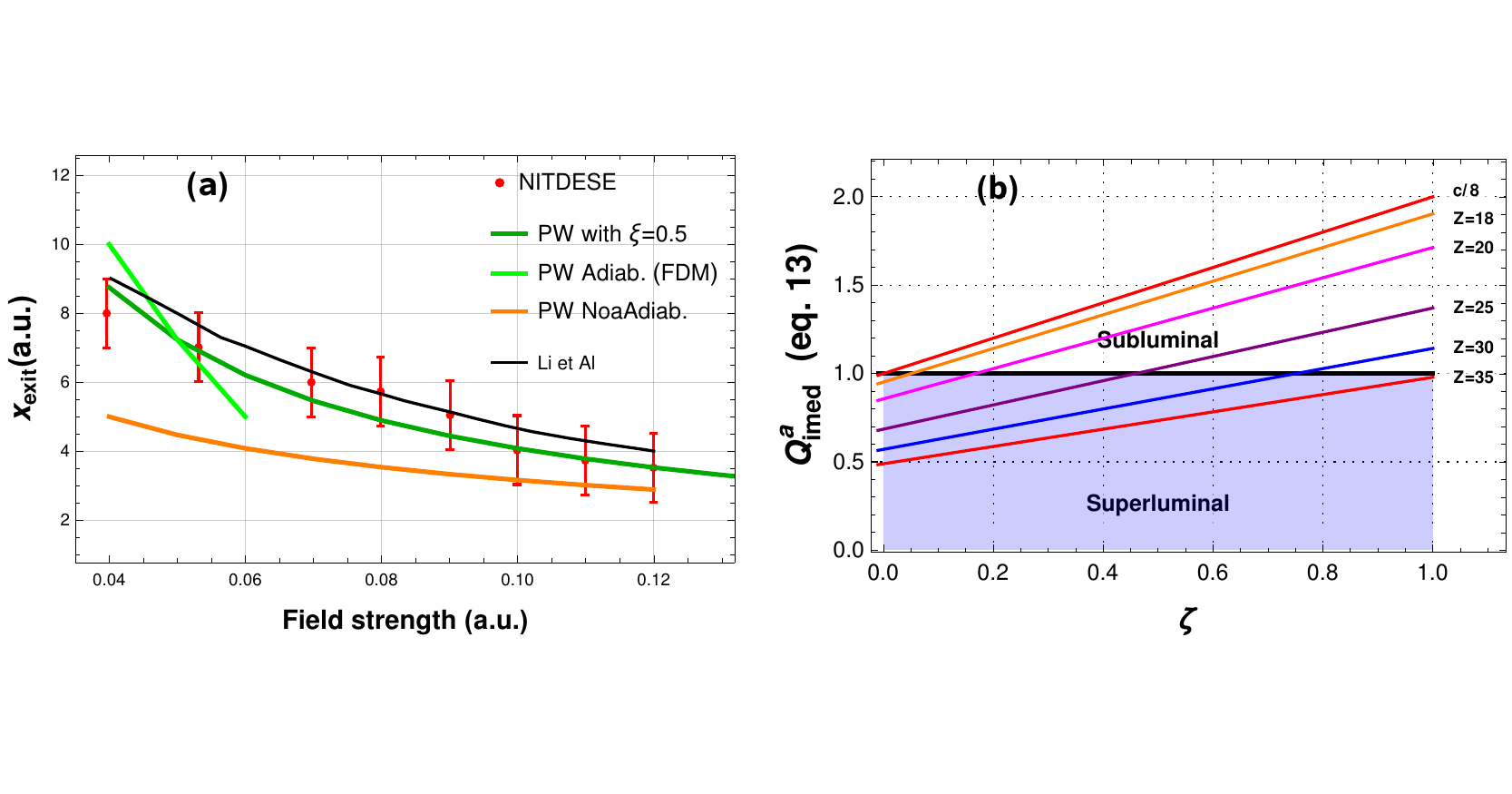}
 \vspace{-1.20cm}
 \caption{\label{figQSbtdzeta} (Color online) 
 (a): The plot shows the exit point (in au) of different approaches
for hydrogen atom versus field strength F (in au), see Figs. \ref{figptc}, \ref{figTI}, sec. \ref{sec:int} and Eq. \ref{timed1}. 
 (b): $Q^{a}_{imed}$ (Eq. \ref{qsimeda}) vs. $\zeta$ for 
 selected $Z$-values showing transition from subluminal to superluminal 
 (see text). }
 \end{figure}
 We illustrate this in Fig. \ref{figQSbtdzeta}~(b), where we plot 
 $Q^{a}_{imed}$ vs. $\zeta$ for selected $Z$ values, which  shows the 
 linear dependence on $\zeta$ (Eq. \ref{qsimeda}).
 As seen in Fig .\ref{figQSbtdzeta}~(b), for a given $Z$ (e.g. $Z_i$) or a segment $(Z_{i},\zeta)$ superluminal tunnel-ionization (colored area) exists as long as $\zeta$ is below a value $\zeta_{i}$, which is  determined by the intersection with the black line $Q^{a}_{imed}=1$, which is given by  $\zeta_{i}=-1+{8 Z_{i}}/{c}$. 
 Or for a chosen $Z_{i}$, the tunnel-ionization  is subliminal when $\zeta>\zeta_{i}$. 
 Conversely, QS (colored area) exists for a given value $\zeta_{i}$ (at an intersection with the black line) for all $Z$ values larger than $Z_{i}$ that corresponds to $\zeta_{i}$.
 This can be seen for a chosen $\zeta_{i}$ (on the black line)  through a vertical line at $\zeta_{i}$ on the $\zeta$-axis in Fig.~\ref{figQSbtdzeta}~(b).  
 At $\zeta=1$ (or $\zeta_{i}=\zeta_{35}$) the QS satisfies the condition of Eq. \ref{qsad} (i.e. the adiabatic tunnel-ionization is superluminal for all $Z\ge 35$) as clearly seen in Fig. \ref{figQSbtdzeta}~(b) (colored area) and already discussed in sec. \ref{pa:adb}. 
 While on the opposite side, i.e. $\zeta=0$, QS is only possible for $Z\ge 18\approx c/8$, which corresponds to the nonadiabatic tunnel ionization case described in \ref{pa:nada}.
 
 However, the superluminality discussed here is based on our probably erroneous definition through the use of $\tau^{Ad}_{c}$, which is justified only for the adiabatic case with $\zeta =1$.
 See also discussion in in sec. \ref{pa:nada}. 
 
 \paragraph{With $\tau_{c}^{imed}$:}\label{pa:inb}
\vskip10pt
 The comparison in the previous paragraph \ref{pa:ina} is questionable, because the barrier width $d_{B} (\mbox{or } d_{C})$ was used, whereas the corresponding horizontal  distance to the exit point $x_{imed}$ (see Fig. \ref{figTI}) is given  by $d_{imed}$. 
 And $\tau _{c}^{imed}={d_{imed}}/{c}$ is then the time that the light 
 needs to travel (horizontally) the distance $d_{imed}$ (Eq. 
 \ref{timed1}). 
 Therefore, with Eqs. \ref{timed}, \ref{timed1}  we obtain: 
 \begin{eqnarray}\label{qsimedb}\nonumber
  Q_{imed}^{b}(Z,\zeta,F)&=&
  \frac{\tau_{imed}}{\tau^{imed}_{c}}\\ 
 &=&\frac{c\, (I_{p}+\zeta\, \delta_z)}{(8 Z\, F) (1-\zeta)\, x_{m}+ 8 \, \zeta Z\delta_z}\\
 \label{qsimedb1}
&=& \frac{c\, (1+\zeta)}{(8 Z F/I_{p}) (1-\zeta)\, x_{m}+ 8 \, \zeta Z}\\\label{qsimedb2}
  &\approx&\frac{c\, (1+\zeta)}{16\, (1-\zeta)/ x_{m}+ 8\, \zeta Z}
 \end{eqnarray}
 with $\zeta\in[0,1]$, where in the 3nd line the thick barrier limit $\delta_z\approx I_{p}$ (thick-barrier limit) and in the 4th line $I_{p}=Z^{2}/2$ is used. 
 From Eqs.~\ref{qsimedb}-\ref{qsimedb2} it can be verified that the limits 
 $\zeta\to 1, 0$ give the two cases discussed above in sec \ref{pa:adb} 
 and {\ref{pa:nadb}}, with the condition $Q_{imed}^{b}\le 1 \Rightarrow 
 Z\ge c/4\approx 34.26$ for the two limits.   
 In other words, the condition of $Z\ge 35$ for the QS to happen, is 
 the same regardless of the contributions of vertical and horizontal 
 channels.  We note that it can be shown that (where $F>0$) for $0<Z\le c/8$, $\zeta<0$ and for $c/8<\zeta<c/4$, $\zeta<0$ or $\zeta>1$, i.e.  values that lie  outside the range are obtained .

 
 
\begin{figure}[t]
 \vskip-25pt
  \includegraphics[height=5.cm,width=8.50cm]{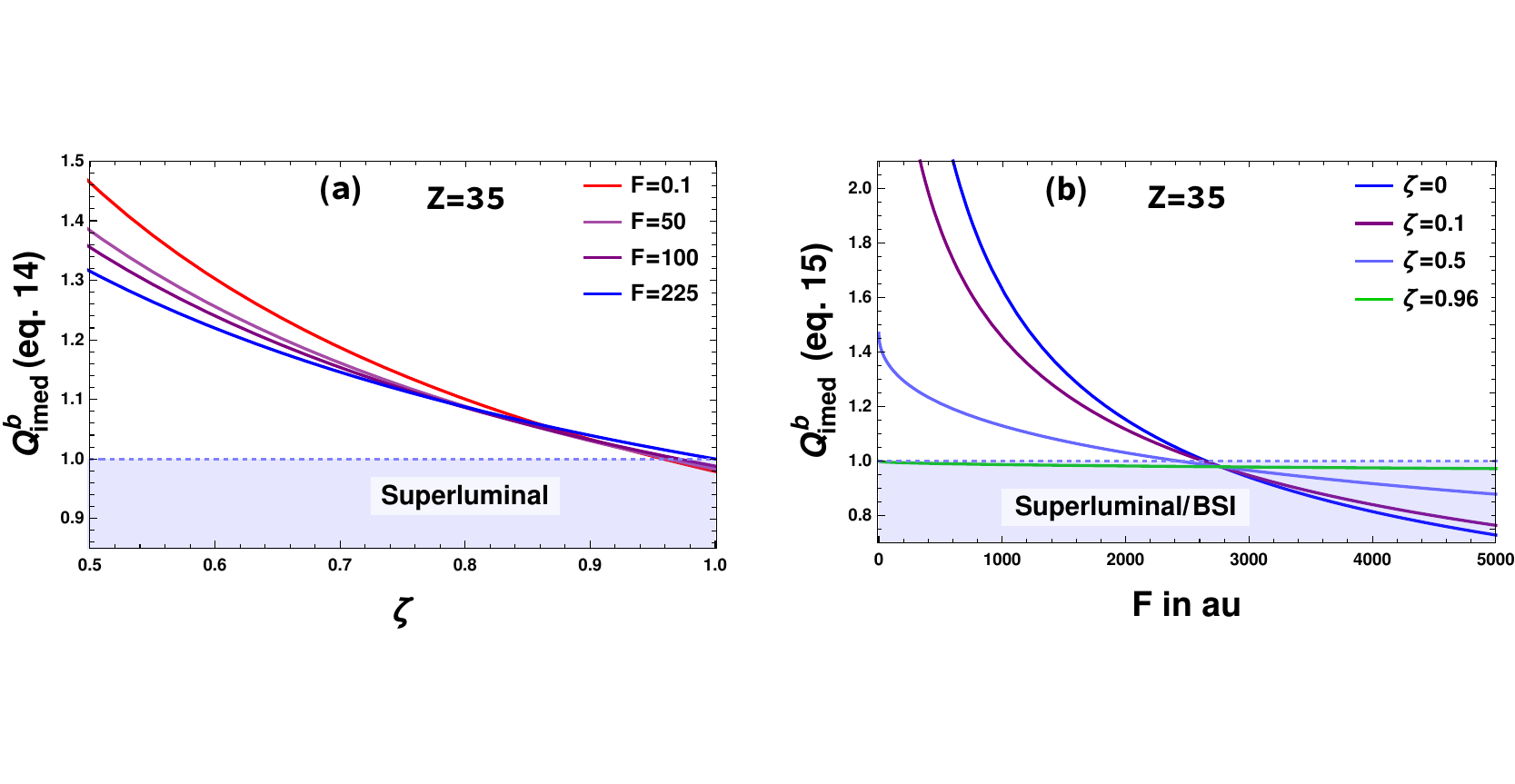}
  \vskip-50pt
  \includegraphics[height=5.cm,width=8.50cm]{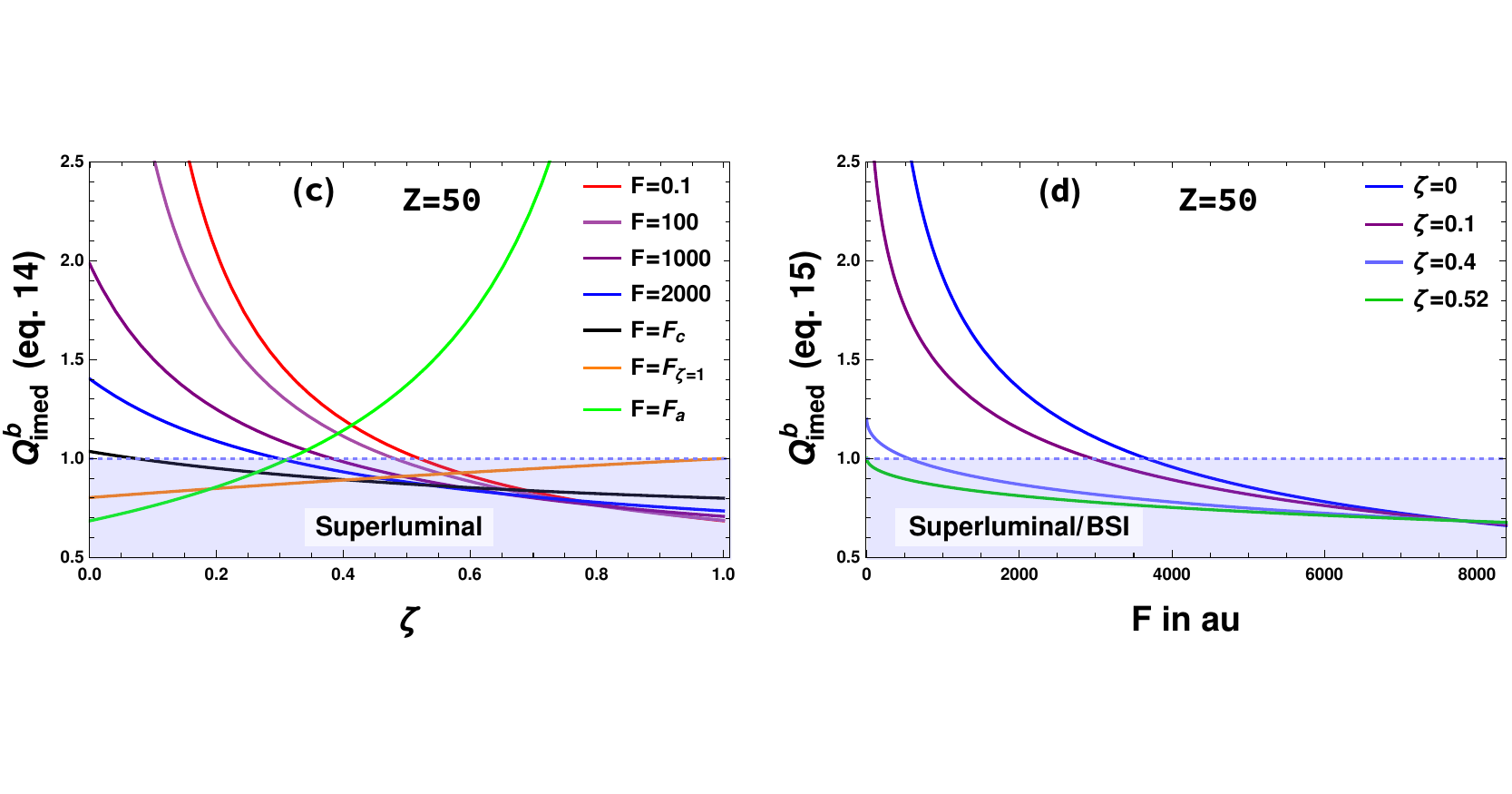}
 \vspace{-1.20cm}
 \caption{\label{figQSbtdzeta1}  
 (Color online) (a),(c): $Q^{b}_{imed}$ of Eq. \ref{qsimedb} vs. $\zeta$ for $Z=35,50$ for different field strengths, showing transition from subluminal to superluminal at the corresponding $\zeta_{Qs}$. In (c): $F_{c}=3667.75\, au$ (Eq. \ref{Fnonad}), $F_{\zeta=1}=6104.5\, au$ and the atomic field strength $F_{a}=8380.3\, au$ (see text).  
 (b),(d) ${Q_{imed}^{b}}$ of Eq. \ref{qsimedb1} vs. field strength for $Z=35,50$ for different $\zeta$ values.} 
 \end{figure}

In Fig. \ref{figQSbtdzeta1}~(a),~(c) $Q_{imed}^{b}$ of the equation \ref{qsimedb} is plotted against $\zeta$ for $Z=35,50$ and various field strengths and in Fig.~\ref{figQSbtdzeta1}~(b),(d) $Q_{imed}^{b}$ of the equation \ref{qsimedb1} against the field strength for $Z=35,50$ and various $\zeta$ values, respectively. We first discuss Fig. \ref{figQSbtdzeta1}~(a),~(c) and then return to Fig. \ref{figQSbtdzeta1}~(b),(d) later.

 In the following, $\zeta_{QS}$ denotes the value of $\zeta$ at the intersection with the dotted line at $Q^{b}_{imed}=1$ in Fig. \ref{figQSbtdzeta1}(a)-(d), which can be determined by solving of the equation $Q^{b}_{imed}(Z,\zeta,F)=1$ of Eq. \ref{qsimedb} or \ref{qsimedb1} with the corresponding $Z$ and $F$ values. 
 
 Considering Fig. \ref{figQSbtdzeta1}~(a),(c) we see that a main feature is that the adiabatic and nonadiabatic contributions (i.e. $\zeta$ value) for which the QS occurs, changes with the field strength. Furthermore, it is evident that for $F\sim 0$ quantum superliminality depends also on $Z$ values.  At the lower value of $Z=35$ (Fig. \ref{figQSbtdzeta1}(a)), QS is only possible for $\zeta_{QS}\sim 1$, which corresponds to the adiabatic or horizontal channel.
 For $Z=35$ this channel is hardly possible at such field strengths (since it is the multiphoton regime because the Keldysh parameter is too large $\gamma_K=\omega \sqrt{2I_{p}}/F\gg 1$, where $\omega$ is the laser frequency and it is of order $1-3\, a.u.$). 
 This means that at field strengths $F\sim 0$, the probability amplitude for such a process is expected to be very small. While $\zeta_{QS}$ changes slowly with increasing $F$ values, the probability amplitude is expected to increase as a result. 
 
 For larger $Z$, as we see in Fig.~\ref{figQSbtdzeta1}(c) for $Z=50$, the situation changes and  QS occurs at smaller $\zeta_{QS}$ values and changes gradually with $F$, where for small field strength we have $\zeta_{QS}(F\sim0)\approx0.52$, the solution of $Q^{b}_{imed}(Z=50,\zeta,F\sim0)=1$.
 At higher field strengths as $\zeta_{QS}(F)$ decreases, the adiabatic contribution decrease, while the multiphoton contribution increases and QS occurs at the corresponding $\zeta_{QS}(F)$ value. 
 Once the  field strength reaches $F_{c}=(c/16)^2 Z$ (compare Eq. \ref{Fnonad}), QS becomes already possible at $\zeta(F_{c})\sim 0$ (and for $\zeta\ge0$, black curve in  Fig.~\ref{figQSbtd1}(c)), where the nonadiabatic tunnel-ionization or the vertical channel is superluminal as we have seen in section \ref{pa:nadb}, and an additional contribution from the horizontal channel is then no longer so important. 
 This trend continues as $F$ increases until $F_{i}=F_{\zeta=1}$, which is the solution of  $Q^{b}_{imed}(Z_{i},\zeta=1,F)=1$ with the corresponding $Z=Z_{i}$ value ($Z_{i}=50$ in our case, orange colored line in Fig.~\ref{figQSbtdzeta1}~(c)).
 
 For larger field strengths $F_{\zeta=1}<F<F_{a}$ the trend turns now in the opposite direction and the vertical channel (nonadiabatic contribution) is gaining in importance and QS occurs at $\zeta_{QS}<1$ values until $F=F_{a}$ with the smallest value $\zeta_{QS}(F_{a})$ (green curve in Fig.~\ref{figQSbtdzeta1}~(c), where $\zeta_{QS}(F_{a})$ is the solution of $Q^{b}_{imed}(Z_{i},\zeta,F_{a})=1$ (in our case of $Z=50$, $\zeta_{QS}(F_{a})\approx 0.31$), where the QS is then possible for $\zeta\le\zeta_{QS}(F_{a})$ at $F=F_{a}$ although $\delta_{z}=0$ (compare Eq.~\ref{qsimedb}). However, for $F>F_{a}$ (BSI) the QS (of $Q^{b}_{imed}$) becomes imaginary but for $\zeta=0$, because $\delta_{z}$ becomes imaginary. 

 The comparison between the two cases in figure \ref{figQSbtdzeta1}~(a),~(c) shows that as $Z$ increases, the $\zeta$ values at which superluminal speed occurs decrease. This can be easily derived from Eq. \ref{qsimedb}; for example, for every $F$, $\zeta(Z_{2},F)<\zeta(Z_{1},F)$ is obtained if $Z_{2}>Z_{1}$. The higher $Z$ is, the smaller the $\zeta$ value at which the intersection with $Q^{b}_{imed}=1$ occurs (as can be seen from Eq.~\ref{qsimedb}-\ref{qsimedb2} by setting $Q^{b}_{imed}=1$). Otherwise, as $Z$ increases, the behavior for all $Z>35$ approaches that of the value $Z=50$ in Fig.~\ref{figQSbtdzeta1}~(c) and remains similar for all large  $Z>35$.
 
 It worth noting that the above discussed behavior of tunnel-ionization actually differs from known studies on tunneling (usually on system with small $Z$ or $Z_{eff}$), where adiabatic and nonadiabatic channel are even studied experimentally separately, see e.g. \cite{Kullie:2025} and the references therein, referred to as adiabatic and nonadiabatic field calibration.
 Therefore, we think that the approximations in Eqs.~\ref{timed},\ref{timed1} suggest to investigate tunnel-ionization mechanism, where the two channels  contribute to the tunneling, compare Fig.~\ref{figQSbtdzeta}~(a).
  In fact, this has already been discussed in earlier works on tunneling mechanism, e.g. by Ivanov et al.~\cite{Ivanov:2005}, who proposed the coexistence of horizontal and vertical channels, as outlined in Fig. \ref{figTI} (intermediate curves). 
   
  

 We now turn to Fig.~\ref{figQSbtdzeta1}~(b),~(d), where we plotted Eq.~\ref{qsimedb1} (where the approximation $\delta_z=I_{p}$ and $d_{B}\approx d_{C}$ are used) against the field strength $F$ for various $\zeta$ values. As can be seen from the figures, similar features are observed as in Fig.~\ref{figQSbtdzeta1}~(a),~(c). For smaller $Z$, i.e. $Z=35$ in Fig.~\ref{figQSbtdzeta1}~(b), for small field strengths superluminal tunnel-ionization only occurs for $\zeta\sim 0.96$ value, i.e. near the adiabatic limit (green curve in Fig.~\ref{figQSbtdzeta1}~(b)). Otherwise it occurs at exclusively large field strengths for all $\zeta$ value smaller that $\zeta_{QS}\approx 0.96$, which is found by solving $Q_{imed}^{b}=1$ of Eq.~\ref{qsimedb1} and taking the limit at small $F$, which changes only slightly with increasing field strength $F$. With further increasing field strength, barrier suppression ionization (BSI) is reached, in which QS also occurs. It should be noted, however, that in Eq.~\ref{qsimedb1} $d_{C}=\nicefrac{I_{p}}{F}$ is used instead of $d_{B}=\nicefrac{\delta_z}{F}$ (see Eq.~\ref{timed1}). 

 For larger $Z$, e.g. $Z=50$ in Fig.~\ref{figQSbtdzeta1}~(d), we observe a similar behavior but $\zeta_{QS}$ becomes smaller with increasing $Z$ value and changes significantly with increasing field strength $F$. At small or realistic field strengths of $F\sim 100-500$, the value is approximately $\zeta_{QS}\sim 0.52$ (green curve in Fig.~\ref{figQSbtdzeta1}~(d)), where the adiabatic and nonadiabatic contributions are sufficiently balanced, compare Fig.~\ref{figQSbtdzeta}~(a). For small $\zeta_{QS}$ values, the probability of superluminal tunnel-ionization can be high, as this corresponds to the multiphoton regime (blue and purple colored curves in Fig.\ref{figQSbtdzeta1}(d)). However, the field strengths needs to be too high for this to occur.
 
Our final result is shown in Fig.~\ref{figQSbtdtau}. There, the tunnel-ionization time-delay $\tau_{imed}$ and the corresponding $\tau_{c}^{imed}$ are plotted as a function of $F$ for small and experimentally reliable field strengths for selected $Z$ values. For clarity in Fig.~\ref{figQSbtdtau}, a slightly larger $\zeta$-value than $\zeta_{QS}$ for the corresponding $Z$ is used, which as already mentioned is determined from Eq.~\ref{qsimedb} by solving the equation $Q^{b}_{imed}=1$ and taking the limit for small $F$ (compare Fig.~\ref{figQSbtdzeta1}(a),(c)). As can be seen in Fig. \ref{figQSbtdtau}, the curves for $\tau_{c}^{imed}$ (green dotted curves) are somewhat higher or above the respective $\tau_{imed}$ (i.e. $\tau_{imed}<\tau^{imed}_{c}$), which becomes more apparent the larger $Z$ is. 
 For the smaller $Z$ values, i.e. $Z=35$, $\zeta_{QS}\approx 0.96$ the tunnel-ionization happens almost adiabatic (horizontal channel). The adiabatic contribution is only slightly below its maximum at $\zeta=1$, however, it is not expected that the probability of tunnel ionization will be significantly increased compared to adiabatic tunnel ionization (corresponding to $\zeta=1$).
 
 As already mentioned (compare Fig. \ref{figQSbtdzeta1}~(a),(c)), a decrease in $\zeta$, although it implies an increase in $Z$, means a decrease in the adiabatic contribution, where one would expect an increase in the tunnel-ionization probability (despite that the barrier becomes thicker and higher) as multiphoton absorption (vertical channel) becomes important. 
 For $Z=50$, we find $\zeta_{QS}\sim 0.52$ for the occurrence of QS at small field strength, so that the two adiabatic/nonadiabatic contributions are in a balanced ratio. This represents a case that appears experimentally feasible, and its experimental investigation and confirmation would be an important step and an insightful finding for tunnel theory in general.
 With a further increase in the $Z$ value, e.g.  $Z =100$ shown in Fig. \ref{figQSbtdtau}, QS  becomes even more pronounced, such that the non-adiabatic contribution predominates (since $\zeta_{QS} \sim 0.2$ is relatively small). However, for such a large $Z$, the barrier becomes too thick at small field strengths, and the probability amplitude for such a process would likely be small.
 \subsection{Summary}
 In summary, our approach is consistent and provides evidence for, or clarifies, the QS of tunnel-ionization. 
 Our result represents an important step towards understanding this fundamental physical property of tunnel-ionization; experimental verification using the attoclock scheme is possible, albeit under somewhat strict conditions.
 \begin{figure}[t]
 \includegraphics[width=8.0cm]{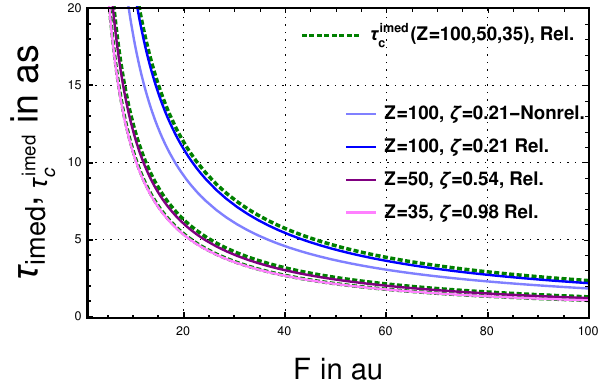}
 \vspace{-0.40cm}
 \caption{\label{figQSbtdtau}  
 (Color online) 
 Time-delay $\tau_{imed}$ (Eq.~\ref{timed}) for $Z=35,50,100$ values and the corresponding $\tau_{c}^{imed}=d_{imed}/c$ (green dashed for all $Z$), see Eq.~\ref{timed1} and sec.~\ref{pa:inb}, vs. field strength $F$, showing QS for $Z\ge 35$ for different $(Z, \zeta)$-values.} 
 \end{figure}
 We summarize our result for H-like atomic ions  as follows: 
 \vspace{-0.2cm}  
 \begin{enumerate}
 \item In the adiabatic case, sec \ref{pa:ad} (see Fig.~\ref{figTI} red 
 dashed-dotted curve), the QS is a fundamental property of adiabatic 
 tunnel-ionization. We found that the barrier time-delay (compare Eqs.~\ref{tdb},~\ref{zvalue} and Fig.~\ref{figQSbtd}) exhibits QS if $Z\ge 18$ at field strengths $F\le F_{a}$. 
 However, the probability for such process 
 is most likely to be low~\cite{Dumont:2023,Winful:2006II,Lunardi:2019}.
 \item In the nonadiabatic case, sec. \ref{pa:nad} (see Fig. \ref{figTI} 
 (dashed light blue curve), we found that the tunnel-ionization or interaction time-delay is superluminal under the condition $Z\ge 35,\,F\ge F_{c}=(\nicefrac{c}{16}) ^{2}Z$ when $\tau_{c}^{Nad}$ is correctly defined by using the correct barrier width, which corresponds to $d_{m}\approx x_{m}$, as shown in~\ref{pa:nadb}. 
 \item The intermediate case, sec.~\ref{pa:in} (see Fig.~\ref{figTI} magenta and purple dotted curves), is important from both fundamental and experimental (or application) point of view. It actually represents the general case, which includes also the two limiting cases, the adiabatic and nonadiabatic case.  We found in sec.~\ref{pa:inb} that under the relatively stringent condition $Z\ge 35, \, 0\le\zeta\le1$ (compare Eq. \ref{qsimedb}, \ref{qsimedb1}), the QS occurs with a varying adiabatic/nonadiabatic contribution dependent on $\zeta$. However, $\zeta$ becomes small for larger $Z$ (e.g., $Z=50$ in Fig.~\ref{figQSbtdzeta1}(c),(d)). Since this implies a decrease in the ratio of adiabatic to non-adiabatic (multiphoton) contribution, an increase in the tunnel-ionization probability is to be expected. 
 \item Our final result is shown in Fig.~\ref{figQSbtdtau}. It clearly demonstrate the possibility of QS in tunnel-ionization.  
 Although the conditions are somewhat strict, the result can be experimentally investigated using the attoclock framework, especially that the field strengths for such a process are available.
 \end{enumerate}
 The probability  or the electron momentum distribution of tunnel-ionization requires a separate investigation in all of these cases. 
 Furthermore, an important result is found in sec.~\ref{pa:ad}, where the QS condition for the barrier time $\tau_{dB}$ of Eq.~\ref{tdb} is less severe ($Z\ge 18$), as shown 
 in Eq.~\ref{zvalue} and Fig.~\ref{figQSbtd}. 
 Although $\tau_{dB}$ cannot be measured directly, it can be extracted from the two calibrations (adiabatic and nonadiabatic) of the same experimental result, in a similar way to the experimental result of the He atom~\cite{Landsman:2014II,Hofmann:2019} as shown in our previous work, compare fig 3 in~\cite{Kullie:2025}.  
 The barrier time-delay $\tau_{dB}$ corresponds to the well known dwell 
 time in the framework of  UTTP \cite{Winful:2003,Lunardi:2019,Kullie:2025}, 
 and to the LC time-delay and the interaction time (in the weak measurement
 and thick-barrier limit)
 \cite{Buettiker:1983,Steinberg:1995,Ramos:2020,Spierings:2021}.
 The thick-barrier limit $\delta\approx I_p$   
 have a negligible effect in our approach because 
 $I_{p}\sqrt{1-{4 F Z}/{I_{p}}^{2}} \approx I_{p}$ for large $Z$ and 
 small field strength. 
 Experimentally, verifying these results using the attoclock scheme is  
 straightforward, as the required field strengths are available. 
 Carrying out the experiment on H-like atoms with large $Z$ could be difficult, but we think it does not represent a major obstacle.
\paragraph*{\bf Acknowledgments}
 I would like to thank Prof. Martin Garcia from the Theoretical
 Physics of the Institute of Physics at the University of Kassel
 for his kind support.

\begin{thebibliography}{42}%
\makeatletter
\providecommand \@ifxundefined [1]{%
 \@ifx{#1\undefined}
}%
\providecommand \@ifnum [1]{%
 \ifnum #1\expandafter \@firstoftwo
 \else \expandafter \@secondoftwo
 \fi
}%
\providecommand \@ifx [1]{%
 \ifx #1\expandafter \@firstoftwo
 \else \expandafter \@secondoftwo
 \fi
}%
\providecommand \natexlab [1]{#1}%
\providecommand \enquote  [1]{``#1''}%
\providecommand \bibnamefont  [1]{#1}%
\providecommand \bibfnamefont [1]{#1}%
\providecommand \citenamefont [1]{#1}%
\providecommand \href@noop [0]{\@secondoftwo}%
\providecommand \href [0]{\begingroup \@sanitize@url \@href}%
\providecommand \@href[1]{\@@startlink{#1}\@@href}%
\providecommand \@@href[1]{\endgroup#1\@@endlink}%
\providecommand \@sanitize@url [0]{\catcode `\\12\catcode `\$12\catcode
  `\&12\catcode `\#12\catcode `\^12\catcode `\_12\catcode `\%12\relax}%
\providecommand \@@startlink[1]{}%
\providecommand \@@endlink[0]{}%
\providecommand \url  [0]{\begingroup\@sanitize@url \@url }%
\providecommand \@url [1]{\endgroup\@href {#1}{\urlprefix }}%
\providecommand \urlprefix  [0]{URL }%
\providecommand \Eprint [0]{\href }%
\providecommand \doibase [0]{http://dx.doi.org/}%
\providecommand \selectlanguage [0]{\@gobble}%
\providecommand \bibinfo  [0]{\@secondoftwo}%
\providecommand \bibfield  [0]{\@secondoftwo}%
\providecommand \translation [1]{[#1]}%
\providecommand \BibitemOpen [0]{}%
\providecommand \bibitemStop [0]{}%
\providecommand \bibitemNoStop [0]{.\EOS\space}%
\providecommand \EOS [0]{\spacefactor3000\relax}%
\providecommand \BibitemShut  [1]{\csname bibitem#1\endcsname}%
\let\auto@bib@innerbib\@empty
\bibitem [{\citenamefont {Kullie}(2015)}]{Kullie:2015}%
  \BibitemOpen
  \bibfield  {author} {\bibinfo {author} {\bibfnamefont {O.}~\bibnamefont
  {Kullie}},\ }\href {\doibase 10.1103/PhysRevA.92.052118} {\bibfield
  {journal} {\bibinfo  {journal} {Phys. Rev. A}\ }\textbf {\bibinfo {volume}
  {92}},\ \bibinfo {pages} {052118} (\bibinfo {year} {2015})},\ \bibinfo {note}
  {\rm arXiv:1505.03400v2}\BibitemShut {NoStop}%
\bibitem [{\citenamefont {Kullie}\ and\ \citenamefont
  {Ivanov}(2024)}]{Kullie:2024}%
  \BibitemOpen
  \bibfield  {author} {\bibinfo {author} {\bibfnamefont {O.}~\bibnamefont
  {Kullie}}\ and\ \bibinfo {author} {\bibfnamefont {I.~A.}\ \bibnamefont
  {Ivanov}},\ }\href {\doibase https://doi.org/10.1016/j.aop.2024.169648}
  {\bibfield  {journal} {\bibinfo  {journal} {Annals of Physics}\ }\textbf
  {\bibinfo {volume} {464}},\ \bibinfo {pages} {169648} (\bibinfo {year}
  {2024})}\BibitemShut {NoStop}%
\bibitem [{\citenamefont {Landsman}\ \emph {et~al.}(2014)\citenamefont
  {Landsman}, \citenamefont {Weger}, \citenamefont {Maurer}, \citenamefont
  {Boge}, \citenamefont {Ludwig}, \citenamefont {Heuser}, \citenamefont
  {Cirelli}, \citenamefont {Gallmann},\ and\ \citenamefont
  {Keller}}]{Landsman:2014II}%
  \BibitemOpen
  \bibfield  {author} {\bibinfo {author} {\bibfnamefont {A.~S.}\ \bibnamefont
  {Landsman}}, \bibinfo {author} {\bibfnamefont {M.}~\bibnamefont {Weger}},
  \bibinfo {author} {\bibfnamefont {J.}~\bibnamefont {Maurer}}, \bibinfo
  {author} {\bibfnamefont {R.}~\bibnamefont {Boge}}, \bibinfo {author}
  {\bibfnamefont {A.}~\bibnamefont {Ludwig}}, \bibinfo {author} {\bibfnamefont
  {S.}~\bibnamefont {Heuser}}, \bibinfo {author} {\bibfnamefont
  {C.}~\bibnamefont {Cirelli}}, \bibinfo {author} {\bibfnamefont
  {L.}~\bibnamefont {Gallmann}}, \ and\ \bibinfo {author} {\bibfnamefont
  {U.}~\bibnamefont {Keller}},\ }\href {\doibase
  http://dx.doi.org/10.1364/OPTICA.1.000343} {\bibfield  {journal} {\bibinfo
  {journal} {Optica}\ }\textbf {\bibinfo {volume} {1}},\ \bibinfo {pages} {343}
  (\bibinfo {year} {2014})}\BibitemShut {NoStop}%
\bibitem [{\citenamefont {Hofmann}\ \emph {et~al.}(2019)\citenamefont
  {Hofmann}, \citenamefont {Landsman},\ and\ \citenamefont
  {Keller}}]{Hofmann:2019}%
  \BibitemOpen
  \bibfield  {author} {\bibinfo {author} {\bibfnamefont {C.}~\bibnamefont
  {Hofmann}}, \bibinfo {author} {\bibfnamefont {A.~S.}\ \bibnamefont
  {Landsman}}, \ and\ \bibinfo {author} {\bibfnamefont {U.}~\bibnamefont
  {Keller}},\ }\href {\doibase 10.1080/09500340.2019.1596325} {\bibfield
  {journal} {\bibinfo  {journal} {J. Mod. Opt.}\ }\textbf {\bibinfo {volume}
  {66}},\ \bibinfo {pages} {1052} (\bibinfo {year} {2019})}\BibitemShut
  {NoStop}%
\bibitem [{\citenamefont {Augst}\ \emph {et~al.}(1989)\citenamefont {Augst},
  \citenamefont {Strickland}, \citenamefont {Meyerhofer}, \citenamefont
  {Chin},\ and\ \citenamefont {Eberly}}]{Augst:1989}%
  \BibitemOpen
  \bibfield  {author} {\bibinfo {author} {\bibfnamefont {S.}~\bibnamefont
  {Augst}}, \bibinfo {author} {\bibfnamefont {D.}~\bibnamefont {Strickland}},
  \bibinfo {author} {\bibfnamefont {D.~D.}\ \bibnamefont {Meyerhofer}},
  \bibinfo {author} {\bibfnamefont {S.~L.}\ \bibnamefont {Chin}}, \ and\
  \bibinfo {author} {\bibfnamefont {J.~H.}\ \bibnamefont {Eberly}},\ }\href
  {\doibase https://doi.org/10.1103/PhysRevLett.63.2212} {\bibfield  {journal}
  {\bibinfo  {journal} {Phys. Rev. Lett.}\ }\textbf {\bibinfo {volume} {63}},\
  \bibinfo {pages} {2212} (\bibinfo {year} {1989})}\BibitemShut {NoStop}%
\bibitem [{\citenamefont {Augst}\ \emph {et~al.}(1991)\citenamefont {Augst},
  \citenamefont {Meyerhofer}, \citenamefont {Strickland},\ and\ \citenamefont
  {Chin}}]{Augst:1991}%
  \BibitemOpen
  \bibfield  {author} {\bibinfo {author} {\bibfnamefont {S.}~\bibnamefont
  {Augst}}, \bibinfo {author} {\bibfnamefont {D.~D.}\ \bibnamefont
  {Meyerhofer}}, \bibinfo {author} {\bibfnamefont {D.}~\bibnamefont
  {Strickland}}, \ and\ \bibinfo {author} {\bibfnamefont {S.~L.}\ \bibnamefont
  {Chin}},\ }\href {\doibase https://doi.org/10.1364/JOSAB.8.000858} {\bibfield
   {journal} {\bibinfo  {journal} {J. Opt. Soc. Am. B}\ }\textbf {\bibinfo
  {volume} {8}},\ \bibinfo {pages} {858} (\bibinfo {year} {1991})}\BibitemShut
  {NoStop}%
\bibitem [{\citenamefont {Delone}\ and\ \citenamefont
  {Kra\v{i}nov}(1998)}]{Delone:1998}%
  \BibitemOpen
  \bibfield  {author} {\bibinfo {author} {\bibfnamefont {N.~B.}\ \bibnamefont
  {Delone}}\ and\ \bibinfo {author} {\bibfnamefont {V.~P.}\ \bibnamefont
  {Kra\v{i}nov}},\ }\href {\doibase
  https://doi.org/10.1070/pu1998v041n05abeh000393} {\bibfield  {journal}
  {\bibinfo  {journal} {Phys.-Usp.}\ }\textbf {\bibinfo {volume} {41}},\
  \bibinfo {pages} {469} (\bibinfo {year} {1998})}\BibitemShut {NoStop}%
\bibitem [{\citenamefont {Kiyan}\ and\ \citenamefont
  {Kra\v{i}nov}(1991)}]{Kiyan:1991}%
  \BibitemOpen
  \bibfield  {author} {\bibinfo {author} {\bibfnamefont {I.~Y.}\ \bibnamefont
  {Kiyan}}\ and\ \bibinfo {author} {\bibfnamefont {V.~P.}\ \bibnamefont
  {Kra\v{i}nov}},\ }\href {http://jetp.ras.ru/cgi-bin/dn/e_073_03_0429.pdf}
  {\bibfield  {journal} {\bibinfo  {journal} {Soviet Phys. JETP}\ }\textbf
  {\bibinfo {volume} {73}},\ \bibinfo {pages} {429} (\bibinfo {year}
  {1991})}\BibitemShut {NoStop}%
\bibitem [{\citenamefont {Kullie}(2016)}]{Kullie:2016}%
  \BibitemOpen
  \bibfield  {author} {\bibinfo {author} {\bibfnamefont {O.}~\bibnamefont
  {Kullie}},\ }\href {\doibase https://doi.org/10.1088/0953-4075/49/9/095601}
  {\bibfield  {journal} {\bibinfo  {journal} {Journal of Physics B: Atomic,
  Molecular and Optical Physics}\ }\textbf {\bibinfo {volume} {49}},\ \bibinfo
  {pages} {095601} (\bibinfo {year} {2016})}\BibitemShut {NoStop}%
\bibitem [{\citenamefont {Kullie}(2018)}]{Kullie:2018}%
  \BibitemOpen
  \bibfield  {author} {\bibinfo {author} {\bibfnamefont {O.}~\bibnamefont
  {Kullie}},\ }\href {\doibase https://doi.org/10.1016/j.aop.2018.01.001}
  {\bibfield  {journal} {\bibinfo  {journal} {Ann. of Phys.}\ }\textbf
  {\bibinfo {volume} {389}},\ \bibinfo {pages} {333} (\bibinfo {year}
  {2018})},\ \bibinfo {note} {\rm arXiv:1701.05012}\BibitemShut {NoStop}%
\bibitem [{\citenamefont {Kullie}(2020)}]{Kullie:2020}%
  \BibitemOpen
  \bibfield  {author} {\bibinfo {author} {\bibfnamefont {O.}~\bibnamefont
  {Kullie}},\ }\href {\doibase https://doi.org/10.3390/quantum2020015}
  {\bibfield  {journal} {\bibinfo  {journal} {Quant. Rep.}\ }\textbf {\bibinfo
  {volume} {2}},\ \bibinfo {pages} {233} (\bibinfo {year} {2020})}\BibitemShut
  {NoStop}%
\bibitem [{\citenamefont {Kullie}(2025)}]{Kullie:2025}%
  \BibitemOpen
  \bibfield  {author} {\bibinfo {author} {\bibfnamefont {O.}~\bibnamefont
  {Kullie}},\ }\href {\doibase 10.1088/2399-6528/adb09c} {\bibfield  {journal}
  {\bibinfo  {journal} {Journal of Physics Communications}\ }\textbf {\bibinfo
  {volume} {9}},\ \bibinfo {pages} {015003} (\bibinfo {year}
  {2025})}\BibitemShut {NoStop}%
\bibitem [{\citenamefont {Winful}(2003{\natexlab{a}})}]{Winful:2003}%
  \BibitemOpen
  \bibfield  {author} {\bibinfo {author} {\bibfnamefont {H.~G.}\ \bibnamefont
  {Winful}},\ }\href {\doibase 10.1103/PhysRevLett.91.260401} {\bibfield
  {journal} {\bibinfo  {journal} {Phys. Rev. Lett.}\ }\textbf {\bibinfo
  {volume} {91}},\ \bibinfo {pages} {260401} (\bibinfo {year}
  {2003}{\natexlab{a}})}\BibitemShut {NoStop}%
\bibitem [{\citenamefont {Lunardi}\ and\ \citenamefont
  {Manzoni}(2019)}]{Lunardi:2019}%
  \BibitemOpen
  \bibfield  {author} {\bibinfo {author} {\bibfnamefont {J.~T.}\ \bibnamefont
  {Lunardi}}\ and\ \bibinfo {author} {\bibfnamefont {L.~A.}\ \bibnamefont
  {Manzoni}},\ }\href {\doibase 10.1088/1742-6596/1391/1/012112} {\bibfield
  {journal} {\bibinfo  {journal} {J. Phys.: Conference Series}\ }\textbf
  {\bibinfo {volume} {1391}},\ \bibinfo {pages} {012112} (\bibinfo {year}
  {2019})}\BibitemShut {NoStop}%
\bibitem [{\citenamefont {Hartman}(1962)}]{Hartman:1962}%
  \BibitemOpen
  \bibfield  {author} {\bibinfo {author} {\bibfnamefont {T.~E.}\ \bibnamefont
  {Hartman}},\ }\href {\doibase 10.1063/1.1702424} {\bibfield  {journal}
  {\bibinfo  {journal} {J. Appl. Phys.}\ }\textbf {\bibinfo {volume} {33}},\
  \bibinfo {pages} {3427} (\bibinfo {year} {1962})}\BibitemShut {NoStop}%
\bibitem [{\citenamefont {Balcou}\ and\ \citenamefont
  {Dutriaux}(1997)}]{Balcou:1997}%
  \BibitemOpen
  \bibfield  {author} {\bibinfo {author} {\bibfnamefont {P.}~\bibnamefont
  {Balcou}}\ and\ \bibinfo {author} {\bibfnamefont {L.}~\bibnamefont
  {Dutriaux}},\ }\href {\doibase 10.1103/PhysRevLett.78.851} {\bibfield
  {journal} {\bibinfo  {journal} {Phys. Rev. Lett.}\ }\textbf {\bibinfo
  {volume} {78}},\ \bibinfo {pages} {851} (\bibinfo {year} {1997})}\BibitemShut
  {NoStop}%
\bibitem [{\citenamefont {Heitmann}\ and\ \citenamefont
  {Nimtz}(1994)}]{Heitmann:1994}%
  \BibitemOpen
  \bibfield  {author} {\bibinfo {author} {\bibfnamefont {W.}~\bibnamefont
  {Heitmann}}\ and\ \bibinfo {author} {\bibfnamefont {G.}~\bibnamefont
  {Nimtz}},\ }\href {\doibase https://doi.org/10.1016/0375-9601(94)91063-4}
  {\bibfield  {journal} {\bibinfo  {journal} {Physics Letters A}\ }\textbf
  {\bibinfo {volume} {196}},\ \bibinfo {pages} {154} (\bibinfo {year}
  {1994})}\BibitemShut {NoStop}%
\bibitem [{\citenamefont {Nimtz}\ and\ \citenamefont
  {Stahlhofen}(2008)}]{Nimtz:2008}%
  \BibitemOpen
  \bibfield  {author} {\bibinfo {author} {\bibfnamefont {G.}~\bibnamefont
  {Nimtz}}\ and\ \bibinfo {author} {\bibfnamefont {A.}~\bibnamefont
  {Stahlhofen}},\ }\href {\doibase https://doi.org/10.1002/andp.20085200603}
  {\bibfield  {journal} {\bibinfo  {journal} {Annalen der Physik}\ }\textbf
  {\bibinfo {volume} {520}},\ \bibinfo {pages} {374} (\bibinfo {year}
  {2008})},\ \bibinfo {note} {arXiv:0901.3968}\BibitemShut {NoStop}%
\bibitem [{\citenamefont {Dumont}\ and\ \citenamefont
  {Rivlin}(2023)}]{Dumont:2023}%
  \BibitemOpen
  \bibfield  {author} {\bibinfo {author} {\bibfnamefont {R.~S.}\ \bibnamefont
  {Dumont}}\ and\ \bibinfo {author} {\bibfnamefont {T.}~\bibnamefont
  {Rivlin}},\ }\href {\doibase 10.1103/PhysRevA.107.052212} {\bibfield
  {journal} {\bibinfo  {journal} {Phys. Rev. A}\ }\textbf {\bibinfo {volume}
  {107}},\ \bibinfo {pages} {052212} (\bibinfo {year} {2023})}\BibitemShut
  {NoStop}%
\bibitem [{\citenamefont {Chiao}\ and\ \citenamefont
  {Steinberg}(1997)}]{Chiao:1997}%
  \BibitemOpen
  \bibfield  {author} {\bibinfo {author} {\bibfnamefont {R.~Y.}\ \bibnamefont
  {Chiao}}\ and\ \bibinfo {author} {\bibfnamefont {A.~M.}\ \bibnamefont
  {Steinberg}},\ }in\ \href {\doibase
  https://doi.org/10.1016/S0079-6638(08)70341-X} {\emph {\bibinfo {booktitle}
  {Progress in Optics}}},\ Vol.~\bibinfo {volume} {37},\ \bibinfo {editor}
  {edited by\ \bibinfo {editor} {\bibfnamefont {E.}~\bibnamefont {Wolf}}}\
  (\bibinfo  {publisher} {Elsevier Publishing},\ \bibinfo {year} {1997})\ pp.\
  \bibinfo {pages} {345--405}\BibitemShut {NoStop}%
\bibitem [{\citenamefont {Chiao}(1999)}]{Chiao:1999}%
  \BibitemOpen
  \bibfield  {author} {\bibinfo {author} {\bibfnamefont {R.~Y.}\ \bibnamefont
  {Chiao}},\ }\href {\doibase 10.1063/1.57888} {\bibfield  {journal} {\bibinfo
  {journal} {AIP Conference Proceedings}\ }\textbf {\bibinfo {volume} {461}},\
  \bibinfo {pages} {3} (\bibinfo {year} {1999})},\ \bibinfo {note}
  {arXiv.9811019}\BibitemShut {NoStop}%
\bibitem [{\citenamefont {Winful}(2003{\natexlab{b}})}]{Winful:20031}%
  \BibitemOpen
  \bibfield  {author} {\bibinfo {author} {\bibfnamefont {H.~G.}\ \bibnamefont
  {Winful}},\ }\href {\doibase 10.1038/424638a} {\bibfield  {journal} {\bibinfo
   {journal} {Nature}\ }\textbf {\bibinfo {volume} {424}},\ \bibinfo {pages}
  {638} (\bibinfo {year} {2003}{\natexlab{b}})}\BibitemShut {NoStop}%
\bibitem [{\citenamefont {Winful}(2003{\natexlab{c}})}]{Winful:20032}%
  \BibitemOpen
  \bibfield  {author} {\bibinfo {author} {\bibfnamefont {H.~G.}\ \bibnamefont
  {Winful}},\ }\href {\doibase 10.1103/PhysRevLett.90.023901} {\bibfield
  {journal} {\bibinfo  {journal} {Phys. Rev. Lett.}\ }\textbf {\bibinfo
  {volume} {90}},\ \bibinfo {pages} {023901} (\bibinfo {year}
  {2003}{\natexlab{c}})}\BibitemShut {NoStop}%
\bibitem [{\citenamefont {B{\"u}ttiker}\ and\ \citenamefont
  {Washburn}(2003)}]{Buettiker:2003}%
  \BibitemOpen
  \bibfield  {author} {\bibinfo {author} {\bibfnamefont {M.}~\bibnamefont
  {B{\"u}ttiker}}\ and\ \bibinfo {author} {\bibfnamefont {S.}~\bibnamefont
  {Washburn}},\ }\href {\doibase 10.1038/422271a} {\bibfield  {journal}
  {\bibinfo  {journal} {Nature}\ }\textbf {\bibinfo {volume} {422}},\ \bibinfo
  {pages} {271} (\bibinfo {year} {2003})}\BibitemShut {NoStop}%
\bibitem [{\citenamefont {Steinberg}\ \emph {et~al.}(1993)\citenamefont
  {Steinberg}, \citenamefont {Kwiat},\ and\ \citenamefont
  {Chiao}}]{Steinberg:1993}%
  \BibitemOpen
  \bibfield  {author} {\bibinfo {author} {\bibfnamefont {A.~M.}\ \bibnamefont
  {Steinberg}}, \bibinfo {author} {\bibfnamefont {P.~G.}\ \bibnamefont
  {Kwiat}}, \ and\ \bibinfo {author} {\bibfnamefont {R.~Y.}\ \bibnamefont
  {Chiao}},\ }\href {\doibase 10.1103/PhysRevLett.71.708} {\bibfield  {journal}
  {\bibinfo  {journal} {Phys. Rev. Lett.}\ }\textbf {\bibinfo {volume} {71}},\
  \bibinfo {pages} {708} (\bibinfo {year} {1993})}\BibitemShut {NoStop}%
\bibitem [{\citenamefont {MacColl}(1932)}]{MacColl:1932}%
  \BibitemOpen
  \bibfield  {author} {\bibinfo {author} {\bibfnamefont {L.~A.}\ \bibnamefont
  {MacColl}},\ }\href {\doibase 10.1103/PhysRev.40.621} {\bibfield  {journal}
  {\bibinfo  {journal} {Phys. Rev.}\ }\textbf {\bibinfo {volume} {40}},\
  \bibinfo {pages} {621} (\bibinfo {year} {1932})}\BibitemShut {NoStop}%
\bibitem [{\citenamefont {Nanni}(2023)}]{Nanni:2023}%
  \BibitemOpen
  \bibfield  {author} {\bibinfo {author} {\bibfnamefont {L.}~\bibnamefont
  {Nanni}},\ }\href {\doibase 10.1007/s12043-022-02511-y} {\bibfield  {journal}
  {\bibinfo  {journal} {Pramana}\ }\textbf {\bibinfo {volume} {97}},\ \bibinfo
  {pages} {37} (\bibinfo {year} {2023})}\BibitemShut {NoStop}%
\bibitem [{\citenamefont {Lunardi}\ and\ \citenamefont
  {Manzoni}(2018)}]{Lunardi:2018}%
  \BibitemOpen
  \bibfield  {author} {\bibinfo {author} {\bibfnamefont {J.~T.}\ \bibnamefont
  {Lunardi}}\ and\ \bibinfo {author} {\bibfnamefont {L.~A.}\ \bibnamefont
  {Manzoni}},\ }\href {\doibase 10.1155/2018/1372359} {\bibfield  {journal}
  {\bibinfo  {journal} {Advances in High Energy Physics}\ }\textbf {\bibinfo
  {volume} {2018}},\ \bibinfo {pages} {"ID1372359"} (\bibinfo {year}
  {2018})}\BibitemShut {NoStop}%
\bibitem [{\citenamefont {Ivanov}\ \emph {et~al.}(2005)\citenamefont {Ivanov},
  \citenamefont {Spanner},\ and\ \citenamefont {Smirnova}}]{Ivanov:2005}%
  \BibitemOpen
  \bibfield  {author} {\bibinfo {author} {\bibfnamefont {M.~Y.}\ \bibnamefont
  {Ivanov}}, \bibinfo {author} {\bibfnamefont {M.}~\bibnamefont {Spanner}}, \
  and\ \bibinfo {author} {\bibfnamefont {O.}~\bibnamefont {Smirnova}},\ }\href
  {\doibase 10.1080/0950034042000275360} {\bibfield  {journal} {\bibinfo
  {journal} {J. Mod. Opt.}\ }\textbf {\bibinfo {volume} {52}},\ \bibinfo
  {pages} {165} (\bibinfo {year} {2005})}\BibitemShut {NoStop}%
\bibitem [{\citenamefont {Landsman}\ and\ \citenamefont
  {Keller}(2015)}]{Landsman:2015}%
  \BibitemOpen
  \bibfield  {author} {\bibinfo {author} {\bibfnamefont {A.~S.}\ \bibnamefont
  {Landsman}}\ and\ \bibinfo {author} {\bibfnamefont {U.}~\bibnamefont
  {Keller}},\ }\href {\doibase http://dx.doi.org/10.1016/j.physrep.2014.09.002}
  {\bibfield  {journal} {\bibinfo  {journal} {Phys. Rep.}\ }\textbf {\bibinfo
  {volume} {547}},\ \bibinfo {pages} {1} (\bibinfo {year} {2015})}\BibitemShut
  {NoStop}%
\bibitem [{\citenamefont {Clementi}\ and\ \citenamefont
  {Raimondi}(1963)}]{Clementi:1963}%
  \BibitemOpen
  \bibfield  {author} {\bibinfo {author} {\bibfnamefont {E.}~\bibnamefont
  {Clementi}}\ and\ \bibinfo {author} {\bibfnamefont {D.~L.}\ \bibnamefont
  {Raimondi}},\ }\href {\doibase https://doi.org/10.1063/1.1712084} {\bibfield
  {journal} {\bibinfo  {journal} {J. Chem. Phys.}\ }\textbf {\bibinfo {volume}
  {38}},\ \bibinfo {pages} {2686} (\bibinfo {year} {1963})}\BibitemShut
  {NoStop}%
\bibitem [{\citenamefont {Spierings}\ and\ \citenamefont
  {Steinberg}(2021)}]{Spierings:2021}%
  \BibitemOpen
  \bibfield  {author} {\bibinfo {author} {\bibfnamefont {D.~C.}\ \bibnamefont
  {Spierings}}\ and\ \bibinfo {author} {\bibfnamefont {A.~M.}\ \bibnamefont
  {Steinberg}},\ }\href {\doibase 10.1103/PhysRevLett.127.133001} {\bibfield
  {journal} {\bibinfo  {journal} {Phys. Rev. Lett.}\ }\textbf {\bibinfo
  {volume} {127}},\ \bibinfo {pages} {133001} (\bibinfo {year}
  {2021})}\BibitemShut {NoStop}%
\bibitem [{\citenamefont {Winful}(2006)}]{Winful:2006II}%
  \BibitemOpen
  \bibfield  {author} {\bibinfo {author} {\bibfnamefont {H.~G.}\ \bibnamefont
  {Winful}},\ }\href {\doibase https://doi.org/10.1016/j.physrep.2006.09.002}
  {\bibfield  {journal} {\bibinfo  {journal} {Physics Reports}\ }\textbf
  {\bibinfo {volume} {436}},\ \bibinfo {pages} {1} (\bibinfo {year}
  {2006})}\BibitemShut {NoStop}%
\bibitem [{\citenamefont {Esposito}(2001)}]{Esposito:2001}%
  \BibitemOpen
  \bibfield  {author} {\bibinfo {author} {\bibfnamefont {S.}~\bibnamefont
  {Esposito}},\ }\href {\doibase 10.1103/PhysRevE.64.026609} {\bibfield
  {journal} {\bibinfo  {journal} {Phys. Rev. E}\ }\textbf {\bibinfo {volume}
  {64}},\ \bibinfo {pages} {026609} (\bibinfo {year} {2001})}\BibitemShut
  {NoStop}%
\bibitem [{\citenamefont {Ivanov}\ \emph {et~al.}(2025)\citenamefont {Ivanov},
  \citenamefont {Kheifets},\ and\ \citenamefont {Landsman}}]{IAIvanov:2025}%
  \BibitemOpen
  \bibfield  {author} {\bibinfo {author} {\bibfnamefont {I.~A.}\ \bibnamefont
  {Ivanov}}, \bibinfo {author} {\bibfnamefont {A.~S.}\ \bibnamefont
  {Kheifets}}, \ and\ \bibinfo {author} {\bibfnamefont {A.~S.}\ \bibnamefont
  {Landsman}},\ }\href {\doibase 10.1364/JOSAB.569384} {\bibfield  {journal}
  {\bibinfo  {journal} {J. Opt. Soc. Am. B}\ }\textbf {\bibinfo {volume}
  {42}},\ \bibinfo {pages} {1878} (\bibinfo {year} {2025})}\BibitemShut
  {NoStop}%
\bibitem [{\citenamefont {Li}\ \emph {et~al.}(2016)\citenamefont {Li},
  \citenamefont {Geng}, \citenamefont {Han}, \citenamefont {Liu}, \citenamefont
  {Peng}, \citenamefont {Gong},\ and\ \citenamefont {Liu}}]{Li:2016}%
  \BibitemOpen
  \bibfield  {author} {\bibinfo {author} {\bibfnamefont {M.}~\bibnamefont
  {Li}}, \bibinfo {author} {\bibfnamefont {J.-W.}\ \bibnamefont {Geng}},
  \bibinfo {author} {\bibfnamefont {M.}~\bibnamefont {Han}}, \bibinfo {author}
  {\bibfnamefont {M.-M.}\ \bibnamefont {Liu}}, \bibinfo {author} {\bibfnamefont
  {L.-Y.}\ \bibnamefont {Peng}}, \bibinfo {author} {\bibfnamefont
  {Q.}~\bibnamefont {Gong}}, \ and\ \bibinfo {author} {\bibfnamefont
  {Y.}~\bibnamefont {Liu}},\ }\href@noop {} {\bibfield  {journal} {\bibinfo
  {journal} {Phys. Rev. A}\ }\textbf {\bibinfo {volume} {93}},\ \bibinfo
  {pages} {013402} (\bibinfo {year} {2016})}\BibitemShut {NoStop}%
\bibitem [{\citenamefont {B\"uttiker}(1983)}]{Buettiker:1983}%
  \BibitemOpen
  \bibfield  {author} {\bibinfo {author} {\bibfnamefont {M.}~\bibnamefont
  {B\"uttiker}},\ }\href {\doibase 10.1103/PhysRevB.27.6178} {\bibfield
  {journal} {\bibinfo  {journal} {Phys. Rev. B}\ }\textbf {\bibinfo {volume}
  {27}},\ \bibinfo {pages} {6178} (\bibinfo {year} {1983})}\BibitemShut
  {NoStop}%
\bibitem [{\citenamefont {Steinberg}(1995)}]{Steinberg:1995}%
  \BibitemOpen
  \bibfield  {author} {\bibinfo {author} {\bibfnamefont {A.~M.}\ \bibnamefont
  {Steinberg}},\ }\href {\doibase 10.1103/PhysRevLett.74.2405} {\bibfield
  {journal} {\bibinfo  {journal} {Phys. Rev. Lett.}\ }\textbf {\bibinfo
  {volume} {74}},\ \bibinfo {pages} {2405} (\bibinfo {year}
  {1995})}\BibitemShut {NoStop}%
\bibitem [{\citenamefont {Ramos}\ \emph {et~al.}(2020)\citenamefont {Ramos},
  \citenamefont {Spierings}, \citenamefont {Racicot},\ and\ \citenamefont
  {Steinberg}}]{Ramos:2020}%
  \BibitemOpen
  \bibfield  {author} {\bibinfo {author} {\bibfnamefont {R.}~\bibnamefont
  {Ramos}}, \bibinfo {author} {\bibfnamefont {D.}~\bibnamefont {Spierings}},
  \bibinfo {author} {\bibfnamefont {I.}~\bibnamefont {Racicot}}, \ and\
  \bibinfo {author} {\bibfnamefont {A.~M.}\ \bibnamefont {Steinberg}},\ }\href
  {\doibase 10.1038/s41586-020-2490-7} {\bibfield  {journal} {\bibinfo
  {journal} {Nature}\ }\textbf {\bibinfo {volume} {583}},\ \bibinfo {pages}
  {529} (\bibinfo {year} {2020})}\BibitemShut {NoStop}%
\bibitem [{\citenamefont {Ivanov}\ and\ \citenamefont
  {Keifets}(2014)}]{IAIvanov:2014}%
  \BibitemOpen
  \bibfield  {author} {\bibinfo {author} {\bibfnamefont {I.~A.}\ \bibnamefont
  {Ivanov}}\ and\ \bibinfo {author} {\bibfnamefont {A.~S.}\ \bibnamefont
  {Keifets}},\ }\href {\doibase https://doi.org/10.1103/PhysRevA.89.021402}
  {\bibfield  {journal} {\bibinfo  {journal} {Phys. Rev. A}\ }\textbf {\bibinfo
  {volume} {89}},\ \bibinfo {pages} {021402} (\bibinfo {year}
  {2014})}\BibitemShut {NoStop}%
\bibitem [{\citenamefont {Ivanov}\ \emph {et~al.}(2016)\citenamefont {Ivanov},
  \citenamefont {Dubau},\ and\ \citenamefont {Kim}}]{IAIvanov_KTKIM:2016}%
  \BibitemOpen
  \bibfield  {author} {\bibinfo {author} {\bibfnamefont {I.~A.}\ \bibnamefont
  {Ivanov}}, \bibinfo {author} {\bibfnamefont {J.}~\bibnamefont {Dubau}}, \
  and\ \bibinfo {author} {\bibfnamefont {K.~T.}\ \bibnamefont {Kim}},\ }\href
  {\doibase 10.1103/PhysRevA.94.033405} {\bibfield  {journal} {\bibinfo
  {journal} {Phys. Rev. A}\ }\textbf {\bibinfo {volume} {94}},\ \bibinfo
  {pages} {033405} (\bibinfo {year} {2016})}\BibitemShut {NoStop}%
\bibitem [{\citenamefont {Nurhuda}\ and\ \citenamefont
  {Faisal}(1999)}]{Nurhuda:1999}%
  \BibitemOpen
  \bibfield  {author} {\bibinfo {author} {\bibfnamefont {M.}~\bibnamefont
  {Nurhuda}}\ and\ \bibinfo {author} {\bibfnamefont {F.~H.~M.}\ \bibnamefont
  {Faisal}},\ }\href {\doibase 10.1103/PhysRevA.60.3125} {\bibfield  {journal}
  {\bibinfo  {journal} {Phys. Rev. A}\ }\textbf {\bibinfo {volume} {60}},\
  \bibinfo {pages} {3125} (\bibinfo {year} {1999})}\BibitemShut {NoStop}%
\end{thebibliography}
%
\appendix
 \section{}
\subsection{\footnotesize Numerical Integration of the Time-Dependent Schr\"odinger Equation}\label{sec:apdx1}
 \begin{figure}[htp]
 \vspace{-1.5cm}
 \includegraphics[height=22cm,width=18.0cm]{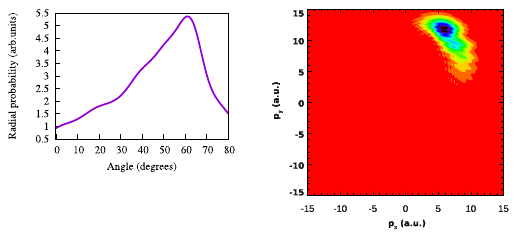}
 \vspace{-18.0cm}
 \caption{\label{figIgor1}   
 (Color online) 
 Photo-electron momentum distribution in the
polarization plane, and radially integrated distribution
defined by eq \ref{rint}. 
Field and target parameters: $Z=18$, $F_{0}=50\, au$, $\omega=3\, au$.}
 \end{figure}
We follow closely the numerical procedure we used to solve the TDSE
in \cite{IAIvanov:2014,IAIvanov_KTKIM:2016}.
We solve the TDSE for a single-electron atom with an effective central
potential:
$\displaystyle V(r)=-\frac{Z_{eff}}{r}$ in the presence of a laser pulse:
\begin{equation}
i \frac{\partial \Psi(\boldsymbol r)} {\partial t}=
\left(\hat H_{\rm atom} + \hat H_{\rm int}(t)\right)
\Psi(\boldsymbol r) \ .
\label{tdse1}
\end{equation}
We use velocity form for the operator $\hat H_{\rm int}(t)$
describing interaction of the atom with the laser field:
\begin{equation}
\hat H_{\rm int}(t) = {\boldsymbol A}(t)\cdot \hat{\boldsymbol p}\ ,
\label{gauge}
\end{equation}
where $\displaystyle \boldsymbol A(t)=-\int\limits_0^t 
\boldsymbol E(\tau)\ d\tau$ is the vector potential of the laser pulse,
which for the geometry we employ (with quantization axis and pulse
propagation direction along the $z$-axis), is defined as follows:
\begin{eqnarray}
A_x(t)&=& -{\frac{f(t)\, F_{0} }{\omega \sqrt{1+\epsilon^2}}}\cos{\omega t}, \, \nonumber \\
A_y(t)&=& {\frac{\epsilon f(t)\, F_{0}} {\omega \sqrt{1+\epsilon^2}}} \sin{\omega t} \ ,
\label{ef}
\end{eqnarray}
where $\epsilon=1$ is ellipticity of the pulse, $F_0$ its field
strength (not to be confused with the {\it peak field strength} $F$
which we use in the formulas in the main text). The function $f(t)$
in eq \ref{ef} is
the pulse envelope which we chose as: $\displaystyle f(t)= \sin^{16}(\pi t/ T_1)$,
where $T_1=2T$, with  $T=2\pi/\omega$- an optical cycle corresponding
to the fundamental frequency $\omega$, is a total duration of the pulse.
The initial state of the system is the ground $1s$ state of an atom with
effective potential $\displaystyle V(r)=-\frac{Z_{eff}}{r}$.
Solution of the TDSE is represented as a  series in spherical harmonics:
\begin{equation}
\Psi({\boldsymbol r},t)=
\sum\limits_{l,m} f_{l}(r,t) Y_l^m(\theta,\phi) \ ,
\label{basis}
\end{equation}
 where spherical harmonics with orders up to $L_{\rm max}=100$ were used
for the field strength $F_0$ we employed in the calculations.
The radial variable is treated by discretizing the TDSE on a grid with
the step-size $\delta r=0.1$ a.u. in a box of the size $R_{\rm max}=400$
a.u. Necessary checks were performed to ensure that for these values
of the parameters $L_{\rm max}$ and $R_{\rm max}$ convergence of the
calculations has been achieved.
The wave-function $\Psi({\boldsymbol r},t)$ was propagated in time
using the matrix iteration method \cite{Nurhuda:1999}.

Ionization amplitude into a photo-electron state with asymptotic
momentum $\boldsymbol p$ is computed by projecting the solution of the
TDSE $\Psi({\boldsymbol r},T_1)$ at the end of the laser pulse on the
scattering states $\phi_{\boldsymbol p}^-$ with ingoing boundary conditions.

We are interested in photo-electron momenta distribution $P(p_x,p_y,0)$
in the polarization $(p_x,p_y)$-plane.
In  Fig. \ref{figIgor1} a typical distribution is shown using the procedure we described above. 
An observable we are after is the offset angle, which for the pulse
defined by eq \ref{ef} is the angle between the negative $y-$
direction and the ray pointing at the maximum of the photo-electron
momentum distribution.
To extract the offset angle, we follow the strategy we employed in
\cite{IAIvanov:2014}).
We compute the radially integrated distribution $P(\phi)$ defined as:
 \begin{equation}
P(\phi)= \int\limits_0^{\infty} P(p_x,p_y,0)p\ dp \ ,
\label{rint}
\end{equation}
where $p=\sqrt{p_x^2+p_y^2}$, factor $p$ under the integral sign in
eq \ref{rint} appears because of the area element in the
$(p_x,p_y)$-plane, and angle $\phi$ is measured from the positive
$x-$ direction.
\subsection{\footnotesize Result}\label{sec:apdx2}
In Fig.\ref{figTI}, the numerical result of the photo-electron momenta 
distribution and the the radially integrated distribution 
$P(\phi)$ for $Z=18$, $F=50\, a.u.$ and $\omega=3\, a.u.$  are shown. 
From the the radially integrated distribution we obtain a time-delay of $\sim 4\, as$. 

 Unfortunately, this value is slightly larger than the corresponding 
 time-delay resulting from our model$\tau_{dion}\sim 1 as$. 
 We do not know at the moment the reason behind this discrepancy. 
 Further calulatios are in prpgress. 
 \end{document}